\documentclass[namedreferences]{solarphysics}
\usepackage[hyperref,optionalrh,solaromanenum]{spr-sola-addons} 
\usepackage{graphicx}                    
\usepackage{color}                       
\usepackage{breakurl}                         
\usepackage{amssymb}
\usepackage{amsbsy}
\usepackage{latexsym}
\usepackage{marvosym}
\usepackage{enumerate}
\usepackage{textcomp}
\usepackage{float}
\usepackage{times}
\usepackage[varg]{txfonts}

\renewcommand{\vec}[1]{{\mathbfit #1}}
\newcommand{\etal}{{\it et al.}}

\begin{document}

\begin{article}

\begin{opening}

\title{Kelvin--Helmholtz instability
       in a cool solar jet in the framework of Hall
       magnetohydrodynamics: A case study}

%
\author[addressref={aff1},corref,email={izh@phys.uni-sofia.bg}]{\inits{I.~}\fnm{I.~}\lnm{Zhelyazkov}\orcid{orcid.org/0000-0001-6320-7517}}\sep
\author[addressref={aff2}]{\inits{Z.~}\fnm{Z.~}\lnm{Dimitrov}}

%
\runningauthor{I.~Zhelyazkov and Z.~Dimitrov}
\runningtitle{Kelvin--Helmholtz instability in a cool solar jet}

\address[id={aff1}]{Faculty of Physics, Sofia University, 1164 Sofia, Bulgaria}
\address[id={aff2}]{Programista JSC, 1407 Sofia, Bulgaria}

\begin{abstract}
We investigate the conditions under which the magnetohydrodynamic (MHD) modes in a cylindrical magnetic flux tube moving along its axis become unstable against the Kelvin--Helmholtz (KH) instability.  We use the dispersion relations of MHD modes obtained from the linearized Hall MHD equations for cool (zero beta) plasma by assuming real wave numbers and complex angular wave frequencies/complex wave phase velocities.  The dispersion equations are solved numerically at fixed input parameters and varying values of the ratio  $l_\mathrm{Hall}/a$, where $l_\mathrm{Hall} = c/\omega_\mathrm{pi}$ ($c$ being the speed of light, and $\omega_\mathrm{pi}$ the ion plasma frequency) and $a$ is the flux tube radius.  It is shown that the stability of the MHD modes depends upon four parameters: the density contrast between the flux tube and its environment, the ratio of external and internal magnetic fields, the ratio $l_\mathrm{Hall}/a$, and the value of the Alfv\'en Mach number defined as the ratio of the tube axial velocity to Alfv\'en speed inside the flux tube.  It is found that at high density contrasts, for small values of $l_\mathrm{Hall}/a$, the kink ($m = 1$) mode can become unstable against KH instability at some critical Alfv\'en Mach number (or equivalently at critical flow speed), but a threshold $l_\mathrm{Hall}/a$ can suppress the onset of the KH instability.  At small density contrasts, however, the magnitude of $l_\mathrm{Hall}/a$ does not affect noticeably the condition for instability occurrence---even though it can reduce the critical Alfv\'en Mach number.  It is established that the sausage mode ($m = 0$) is not subject to the KH instability.
\end{abstract}

%

\end{opening}

%
\section{Introduction}
\label{sec:intro}
\citet{lighthill1960}, almost six decades ago, pointed out that for an adequate description of magnetohydrodynamic (MHD) waves and other anisotropic wave motions through a set of MHD equations it is necessary to take into account the term $m_\mathrm{i}(\vec{j} \times \vec{B})/(e\rho)$, in the generalized Ohm law.  That approach is known as Hall magnetohydrodynamics (Hall MHD).  Hall MHD allows one, for example, to describe waves with angular frequencies up to ion cyclotron frequency, $\omega \approx \omega_\mathrm{ci}$.  Since the model still neglects the electron inertia, it is limited to frequencies well below the lower hybrid frequency, $\omega \ll \omega_\mathrm{lh}$.  Physically the Hall term decouples ion and electron motion on ion inertial length scale, $L < l_\mathrm{Hall} = c/\omega_\mathrm{pi}$ (where $c$ is the speed of light and $\omega_\mathrm{pi}$ is the ion plasma frequency).  In this way, the theory of Hall MHD is relevant to plasma dynamics occurring on length scales shorter than $l_\mathrm{Hall}$, and time scales of the order of or shorter than the ion cyclotron period, $t < \omega_\mathrm{ci}^{-1}$ \citep{huba1995}.  Accordingly, the Hall MHD should affect the dispersion characteristics of the various MHD waves (bulk waves, shocks, solitary waves, surface modes in spatially bounded structures like layers and cylindrical magnetic flux tubes) as well as the emergence and development of turbulence \citep{ghosh1997,galtier2000,galtier2006} or magnetic reconnection \citep{fitzpatrick2004,birn2005}.  The modified form of standard MHD waves, in incompressible limit, due to the Hall effect can be found in \citet{sahraoui2007}.  A review for the studies of propagation of waves in bounded MHD plasmas (in slab geometry) in the context of both the standard and the Hall MHD can be found in \citet{zhelyazkov2009} and references therein.  A fluid model for partially ionized plasma with the Hall term has been built by \citet{pandey2008} and recently was used in studying the surface wave propagation in a partially ionized solar plasma slab \citep{pandey2013}.

A Hall term changes not only the wave's propagating characteristics, but also various kinds of instabilities of space plasmas, and in particular the Kelvin--Helmholtz instability (KHI).  As \citet{chandrasekhar1961} has established, this instability arises at the interface of two incompressible plasmas moving with different velocities embedded in a constant magnetic field if the thin velocity shear around the interface exceeds some critical value.  The KHI in its nonlinear stage can develop a series of KH vortices.  \citet{nykyri2004} examined the influence of the Hall term on KHI and reconnection inside KH vortices at the flank boundaries of the magnetosphere.  It was numerically shown that the arising turbulence might become essentially important in the Earth's magnetospheric cusps.  The influence of environmental parameters on mixing and reconnection caused by the KHI at the magnetopause was very recently explored by \citet{leroy2017}.  These authors studied the different configurations in a three-dimensional Hall-MHD setting, where the double mid-latitude reconnection (DMLR) process is triggered by the equatorial roll-ups.  The impact of various parameters on the growth rate of the KHI and thus the efficiency of the DMLR were also evaluated.  According to \citet{leroy2017}, the studied different configurations may have observable signatures that can be identified by space-borne diagnostics.  We note that, in the case of weakly ionized plasmas, the exploration of KHI becomes more complex because weakly ionized plasmas contain both neutral and charged particle fluids with different physical characteristics.  Interactions between the various species can introduce non-ideal effects.  For example, \citet{shadmehri2008} studied the role of KHI arising at the interface between a partially ionized dusty outflow and the surrounding medium.  They have established that the unstable modes are independent from the charge polarity of the dust particles.  Ambipolar dissipation and the Hall effect are two non-ideal effects that can significantly influence the development of the KHI in a medium by changing the plasma dynamics and the evolution of the magnetic field.  \citet{jones2011} have investigated the behavior of the instability in a Hall-dominated and an ambipolar diffusion dominated plasma with the use of suite of fully multifluid magnetohydrodynamic simulations of KHI using the \textsc{hydra} code.  These authors found that, while the linear growth rates of the instability are unaffected by multifluid effects, its nonlinear behavior is essentially changed by the ambipolar diffusion which removes a large part of magnetic energy.  On the other hand, a strong Hall effect introduces a dynamo effect which leads to continuing strong growth of the magnetic field into the nonlinear regime.  The onset of KHI in dense and cool moving magnetic flux tubes surrounded by a hotter and lighter medium have been studied by \citet{martinez2015} on the base of two-fluid partially ionized plasma hydrodynamics.  These authors have shown that the presence of a neutral component in a plasma may contribute to the onset of KHI even for sub-Alfv\'enic longitudinal shear flows.  A fast-to-Alfv\'en mode conversion in a stratified atmosphere of cold plasma mediated by the Hall current have been explored by \citet{cally2015}.  Their analysis is based on a one-fluid Hall MHD approximation, but under the assumption that the plasma is collisionally dominated.  Thus, in that case, the inertia of neutrals affects the oscillations, notably the full mass density $\rho$ appears in the Alfv\'en speed $v_\mathrm{A} = B/\!\!\sqrt{\mu \rho}$, and not as usual the ion density.  We note that their Hall parameter is defined as $\omega/f\omega_\mathrm{ci}$, where $f$ is the ionization fraction equal to $m_\mathrm{i}n_\mathrm{i}/\rho$, with $m_\mathrm{i}$ and $n_\mathrm{i}$ being the mean ion mass and total number density, respectively.  The authors by assuming an ionization fraction $f$ as low as $10^{-4}$ show that the Hall current can couple low-frequency Alfv\'en and fast magnetoacoustic waves via the aforementioned Hall parameter.  The effect of the Hall term on the onset of the KHI in moving solar structures such as the solar wind was investigated in flat and cylindrical geometry in the incompressible plasma approximation \citep{zhelyazkov2009,zhelyazkov2010}.  It has been shown that in cylindrical geometry \citep{zhelyazkov2010} the kink ($m = 1$) mode, being unstable in the framework of the standard incompressible magnetohydrodynamics, becomes stable in the Hall MHD when the ratio $l_\mathrm{Hall}/a$ possesses a value of $0.4$.  It turns out that the sausage mode ($m = 0$) is always stable in both incompressible magnetohydrodynamics.  It is necessary to note, however, that the derivation of the wave dispersion relation in \citet{zhelyazkov2010} was not quite correct---the thermal/plasma pressure term in the momentum equation was neglected which is unacceptable in the incompressible standard or Hall MHD.  Hence, the results of the numerical studies in that article need a reconsideration.  Here, we present a rigorous derivation of the Hall MHD wave dispersion relation in the limit of zero beta (cool) plasma (being akin to, but not identical to that in \citealp{zhelyazkov2010}) and an exploration of KHI for three density contrast values of the cool jet and its environment at a set of magnitudes for the $l_\mathrm{Hall}/a$ parameter.  A distinctive feature of our new derived dispersion equation is that it allows one to study not only the kink ($m = 1$) MHD mode, but also higher ($m \geqslant 2$) modes, thus widening the region of wave's and jet's plasma parameters that control the onset or suppression of KHI in investigated media.  Moreover, here we found that the onset of KHI crucially depends on the plasma density contrast---a relatively low density contrast can stimulate instability occurrence.

The organization of the paper is as follows: In the next section we give the magnetic field topology of a moving cylindrical flux tube modeling the solar wind alongside the basic governing equations and the Hall MHD normal mode dispersion relation.  Section 3 deals with the numerical solutions to the wave dispersion relation and the discussion of obtained results.  The last section summarizes the new findings and comments on future improvements of the Hall MHD studies of KHI in the solar wind.

\section{Geometry, basic equations, and dispersion relation}
\label{sec:model}
\begin{figure}
 \centerline{\includegraphics[width=0.55\textwidth,clip=]{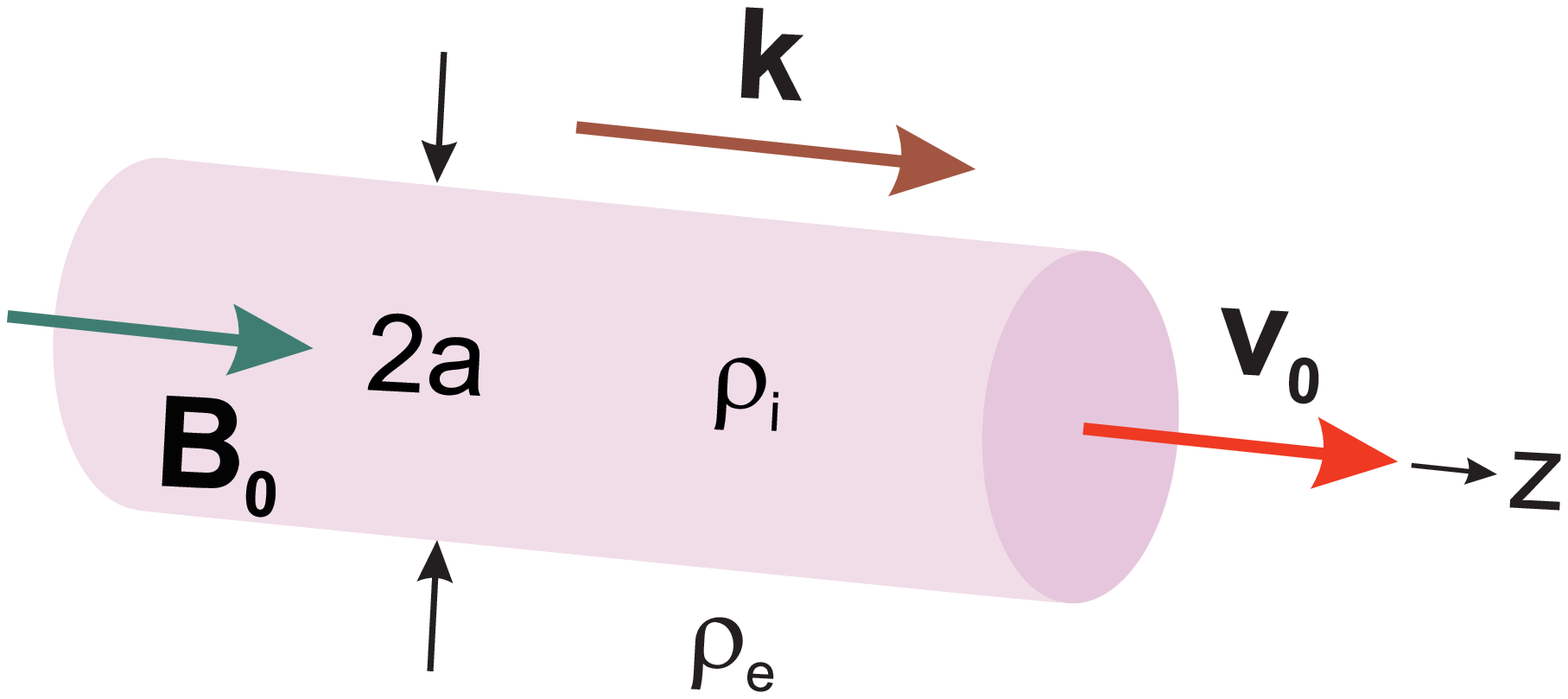}}
 \caption{Magnetic field and velocity configuration in an axially moving solar wind flux tube.}
 \label{fig:fig1}
\end{figure}
We model the solar wind as a moving in axial direction cylindrical magnetic flux tube of cool plasma and radius $a$, embedded in a constant magnetic field $\vec{B}_0 = (0,0,B_0)$ (see Figure~\ref{fig:fig1}).  In the cold-plasma approximation the two magnetic fields (inside and outside the flowing plasma) are identical.  The flux tube velocity $\vec{v}_0$ is homogeneous in radial direction and has only a $z$-component: $\vec{v}_0 = (0,0,v_0)$.  We note that $\vec{v}_0$ is the tube relative velocity with respect to the environment because our frame of reference is attached in the surrounding magnetized plasma.  We assume that plasma densities in both media, $\rho_\mathrm{i}$ and $\rho_\mathrm{e}$, are homogeneous and the ratio $\eta = \rho_\mathrm{e}/\rho_\mathrm{i}$ characterizes the density contrast.  Excited MHD waves propagate along the flux tube, which implies a wavevector $\vec{k} = (0,0,k_z)$.

Considering small perturbations from equilibrium in the form
\[
    \vec{B} = \vec{B}_0 + \vec{B}_1, \quad \rho = \rho_0 + \rho_1, \quad \vec{v} = \vec{v}_0 + \vec{v}_1,
\]
where the subscript $0$ denotes the equilibrium values of magnetic and velocity fields, while the subscript $1$ denotes their perturbations.  The plasma motion is described by the linearized ideal Hall MHD equations for cold plasmas under the consideration that the generalized Ohm law has the form
\begin{equation}
\label{eq:ohm}
    \vec{E} = -\vec{v} \times \vec{B} + \frac{m_\mathrm{i}}{e \rho}\vec{j} \times \vec{B},
\end{equation}
where $e$ is the elementary electric charge, and $m_\mathrm{i}$ is the ion mass.  As is seen, the Hall term yields an additional term to the electric field, which modifies the standard form of the Faraday equation $\partial \vec{B}/\partial t = -\nabla \times \vec{E}$.  With current density $\vec{j} = \mu^{-1} \nabla \times \vec{B}$ (assuming neglected displacement current), the set of linearized equations which govern the dynamics of aforementioned magnetic field and velocity perturbations in the cool-plasma approximation has the form
\begin{equation}
\label{eq:cont}
    \frac{\partial \rho_1}{\partial t} = -\nabla \cdot (\rho_0 \vec{v}_1 + \rho_1 \vec{v}_0) = 0,
\end{equation}
\begin{equation}
\label{eq:momentum}
    \rho_0 \left( \frac{\partial}{\partial t} + \vec{v}_0 \cdot \nabla \right)\vec{v}_1 = \frac{1}{\mu}(\nabla \times \vec{B}_1) \times \vec{B}_0,
\end{equation}
\begin{equation}
\label{eq:faraday}
    \frac{\partial \vec{B}_1}{\partial t} = \nabla \times (\vec{v}_0 \times \vec{B}_1) + \nabla \times (\vec{v}_1 \times \vec{B}_0) - \frac{m_\mathrm{i}}{e \rho \mu}\nabla \times \left[ (\nabla \times \vec{B}_1) \times \vec{B}_0 \right],
\end{equation}
and the constraint
\begin{equation}
\label{eq:divB}
    \nabla \cdot \vec{B}_1 = 0,
\end{equation}
where $\mu$ is the magnetic permeability of free space; the other notation is standard.  Recall that here the quantities with subscript zero refer to equilibrium values.  To investigate the stability of the system, Equations (\ref{eq:momentum})--(\ref{eq:divB}) are Fourier transformed, assuming that all perturbations, in cylindrical coordinates, have the form
\begin{equation}
\label{eq:fourier}
    g(r,\phi,z,t) = g(r)\exp[\mathrm{i}(-\omega t + m \phi + k_z z)],
\end{equation}
where $g$ represents any quantities $\vec{v}_1$ and $\vec{B}_1$; $\omega$ is the angular wave frequency, $m$ is the azimuthal mode number, and $k_z$ is the axial wavenumber.  Equation~(\ref{eq:cont}), which defines the density perturbation, is not used below because we are studying the propagation and stability of Alfv\'en-wave like perturbations of the fluid velocity and magnetic field.  Bearing in mind that for cold plasma in cylindrical coordinate system $\vec{v}_1 = (v_{1r}, v_{1\phi}, 0)$ and $\vec{B}_1 = (B_{1r}, B_{1\phi}, B_{1z})$, if we use Equation~(\ref{eq:fourier}) in Equations (\ref{eq:momentum})--(\ref{eq:divB}), we obtain the following set of equations for the components of the fluid velocity and magnetic field perturbations:
\begin{equation}
\label{eq:v1r}
    -\mathrm{i}\rho_0 \Omega v_{1r} + \frac{\mathrm{d}}{\mathrm{d}r}p_{1\mathrm{m}} - \frac{1}{\mu}B_0 \mathrm{i} k_z B_{1r} = 0,
\end{equation}
\begin{equation}
\label{eq:v1phi}
    -\mathrm{i}\rho_0 \Omega v_{1\phi} + \frac{m}{r}p_{1\mathrm{m}} - \frac{1}{\mu}B_0 k_z B_{1\phi} = 0,
\end{equation}
\begin{equation}
\label{eq:B1r}
    -\mathrm{i}\rho_0 \Omega B_{1r} - B_0 k_z v_{1r} + \mathrm{i}\frac{k_z v_\mathrm{A}^2}{\omega_\mathrm{ci}} \left( \frac{m}{r} B_{1z} - k_z B_{1\phi} \right) = 0,
\end{equation}
\begin{equation}
\label{eq:B1phi}
    -\mathrm{i}\rho_0 \Omega B_{1\phi} - B_0 k_z v_{1\phi} + \mathrm{i}\frac{k_z v_\mathrm{A}^2}{\omega_\mathrm{ci}} \left( k_z B_{1r} + \mathrm{i}\frac{\mathrm{d}}{\mathrm{d}r} B_{1z} \right) = 0,
\end{equation}
\begin{equation}
\label{eq:B1z}
    -\mathrm{i}\rho_0 \Omega B_{1z} + B_0 \left( \frac{\mathrm{d}}{\mathrm{d}r}v_{1r} + \frac{1}{r}v_{1r} + \mathrm{i} \frac{m}{r} v_{1\phi} \right) + \mathrm{i}\frac{k_z v_\mathrm{A}^2}{\omega_\mathrm{ci}}\left( \frac{\mathrm{d}}{\mathrm{d}r}B_{1\phi} + \frac{1}{r}B_{1\phi} - \mathrm{i}\frac{m}{r}B_{1r} \right) = 0,
\end{equation}
where $p_{1\mathrm{m}} = B_0B_{1z}/\mu$ is the magnetic pressure perturbation, $\Omega = \omega - k_z v_0$ is the Doppler shifted frequency, $\omega_\mathrm{ci} = eB_0/m_\mathrm{i}$ is the ion cyclotron frequency, and $v_\mathrm{A} = B_0/\!\!\sqrt{\mu \rho_0}$ is the Alfv\'en speed.  From Equations (\ref{eq:v1r})--(\ref{eq:B1phi}) we obtain the following two coupled equations for $v_{1r}$ and $v_{1\phi}$:
\[
    v_{1r} = \mathrm{i}\frac{\Omega}{k_z^2 v_\mathrm{A}^2 - \Omega^2}\left(\frac{1}{\rho_0}\frac{\mathrm{d}}{\mathrm{d}r} p_{1\mathrm{m}} + \frac{k_z v_\mathrm{A}^2}{\omega_\mathrm{ci}}v_\mathrm{1\phi} \right),
\]
\[
    v_{1\phi} = -\frac{\Omega}{k_z^2 v_\mathrm{A}^2 - \Omega^2}\left(\frac{1}{\rho_0}\frac{m}{r} p_{1\mathrm{m}} + \mathrm{i} \frac{k_z v_\mathrm{A}^2}{\omega_\mathrm{ci}}v_\mathrm{1r} \right),
\]
which allows us to express $v_{1r}$ and $v_{1\phi}$ through the perturbation of the magnetic pressure $p_{1\mathrm{m}}$ and its first derivative with respect to $r$, that is,
\begin{equation}
\label{eq:v1rnew}
    v_{1r} = \mathrm{i}\frac{1}{Z}\frac{\Omega}{k_z^2 v_\mathrm{A}^2 - \Omega^2}\frac{1}{\rho_0}\left( \frac{\mathrm{d}}{\mathrm{d}r}  - \frac{\varepsilon}{1 - C}\frac{m}{r} \right)p_{1\mathrm{m}},
\end{equation}
\begin{equation}
\label{eq:v1phinew}
    v_{1\phi} = -\frac{1}{Z}\frac{\Omega}{k_z^2 v_\mathrm{A}^2 - \Omega^2}\frac{1}{\rho_0}\left( \frac{m}{r} - \frac{\varepsilon}{1 - C}\frac{\mathrm{d}}{\mathrm{d}r} \right)p_{1\mathrm{m}}.
\end{equation}
Here, $Z$, $\varepsilon$, and $C$ are dimensionless expressions that have the forms
\[
    Z = 1 - \left( \frac{\varepsilon}{1 - C} \right)^2, \quad \varepsilon = \Omega/\omega_\mathrm{ci}, \quad \mbox{and} \quad C = \left( \frac{\Omega}{k_z v_\mathrm{A}} \right)^2.
\]

By inserting the above expressions of $v_{1r}$ and $v_{1\phi}$ into Equation~(\ref{eq:B1z}), after some lengthy algebra one obtains
\begin{equation}
\label{eq:A}
    \frac{1}{\rho_0}\frac{1}{k_z^2 v_\mathrm{A}^2 - \Omega^2}\left( 1 - \frac{\varepsilon^2}{1 - C} \right)\left( \frac{\mathrm{d}^2}{\mathrm{d}r^2} + \frac{1}{r}\frac{\mathrm{d}}{\mathrm{d}r} - \frac{m^2}{r^2} \right)p_{1\mathrm{m}} = \frac{B_{1z}}{B_0}.
\end{equation}

From Equation~(\ref{eq:v1phi}) we have
\[
    \frac{m}{r}\frac{1}{\mu}B_0 B_{1z} = \rho_0 \Omega v_{1\phi} + \frac{1}{\mu}B_0 k_z B_{1\phi}.
\]
By expressing $B_{1\phi}$ via the perturbations $v_{1r}$ and $v_{1\phi}$, namely
\[
    B_{1\phi} = - \frac{k_z B_0}{\Omega}\left[ v_{1\phi} + \mathrm{i}(\Omega/\omega_\mathrm{ci})v_{1r} \right],
\]
and multiplying the above equation with $\Omega/\rho_0$, on using Equations~(\ref{eq:v1rnew}) and (\ref{eq:v1phinew}), we obtain 
\begin{equation}
\label{eq:b1zoverb0}
    \frac{B_{1z}}{B_0} = \frac{1}{\rho_0}\frac{1}{v_\mathrm{A}^2}p_{1\mathrm{m}}.
\end{equation}
Then Equation~(\ref{eq:A}) takes the form
\begin{equation}
\label{eq:bessel}
    \left( \frac{\mathrm{d}^2}{\mathrm{d}r^2} + \frac{1}{r}\frac{\mathrm{d}}{\mathrm{d}r} - \frac{m^2}{r^2} - \kappa^2 \right)p_{1\mathrm{m}} =0,
\end{equation}
where
\[
    \kappa^2 = k_z^2\left( 1 - \frac{\Omega^2}{k_z^2 v_\mathrm{A}^2} \right)\frac{1}{Y} \quad \mbox{and} \quad Y = \frac{1 - \varepsilon^2/(1 - C)}{Z}.
\]

Equation~(\ref{eq:bessel}) is the equation for the modified Bessel functions and its solutions in the two media are:
\begin{equation}
\label{eq:p1mag}
    p_{1\mathrm{m}}(r) = \left\{ \begin{array}{lc}
                                 \alpha_\mathrm{i}I_m(\kappa_\mathrm{i}r) & \mbox{for  $\;\,r \leqslant a$} \\
                                 \alpha_\mathrm{e}K_m(\kappa_\mathrm{e}r) & \mbox{for  $\;r > a$},
                                 \end{array}
                         \right.
\end{equation}
where the wave attenuation coefficients $\kappa_\mathrm{i,e}$ are given by the expressions
\[
    \kappa_\mathrm{i} = k_z \left( 1 - \Omega^2/k_z^2 v_\mathrm{Ai}^2 \right)/Y_\mathrm{i} \quad \mbox{and} \quad \kappa_\mathrm{e} = k_z \left( 1 - \omega^2/k_z^2 v_\mathrm{Ae}^2 \right)/Y_\mathrm{e}.
\]

With the above expressions for the magnetic pressure perturbation, the radial component (\ref{eq:v1rnew}) of the velocity perturbation has the following presentations in both media:
\begin{eqnarray}
\label{eq:vrie}
    v_{1r}(r \leqslant a) &=& -\mathrm{i}\frac{1}{\rho_\mathrm{i}}\frac{\Omega}{\Omega^2 - k_z^2 v_\mathrm{Ai}^2} \frac{\alpha_\mathrm{i}}{Z_\mathrm{i}}\left[ \kappa_\mathrm{i}I_m^{\prime}(\kappa_\mathrm{i}r) - \frac{\varepsilon_\mathrm{i}}{1 - C_\mathrm{i}} \frac{m}{r} I_m(\kappa_\mathrm{i}r) \right], \nonumber \\
    \\
    v_{1r}(r > a) &=& -\mathrm{i}\frac{1}{\rho_\mathrm{e}}\frac{\omega}{\omega^2 - k_z^2 v_\mathrm{Ae}^2} \frac{\alpha_\mathrm{e}}{Z_\mathrm{e}}\left[ \kappa_\mathrm{e}K_m^{\prime}(\kappa_\mathrm{e}r) - \frac{\varepsilon_\mathrm{e}}{1 - C_\mathrm{e}} \frac{m}{r} K_m(\kappa_\mathrm{e}r) \right], \nonumber
\end{eqnarray}
where the prime means differentiation of the Bessel function on its argument.  Finally, by applying the boundary conditions for continuity of the ratio $v_{1r}/\Omega$ (\emph{aka\/} the radial component $\xi_r$ of the Lagrangian displacement $\pmb{\xi}$) and the magnetic pressure perturbation $p_{1\mathrm{m}}$ at $r = a$, we obtain the dispersion relation of normal MHD modes propagating on a moving cool-plasma magnetic flux tube
\begin{eqnarray}
\label{eq:dispeqn}
    \frac{\rho_\mathrm{e}}{\rho_\mathrm{i}}\left( \omega^2 - k_z^2 v_\mathrm{Ae}^2 \right)Z_\mathrm{e}\left( \kappa_\mathrm{i} \frac{I_m^{\prime}(\kappa_\mathrm{i}a)}{I_m(\kappa_\mathrm{i}a)} - \frac{\varepsilon_\mathrm{i}}{1 - C_\mathrm{i}} \frac{m}{a} \right) \nonumber \\
    \\
    {}-\left[ \left( \omega - k_z v_0 \right)^2 - k_z^2 v_\mathrm{Ai}^2 \right]Z_\mathrm{i}\left( \kappa_\mathrm{e} \frac{K_m^{\prime}(\kappa_\mathrm{e}a)}{K_m(\kappa_\mathrm{e}a)} - \frac{\varepsilon_\mathrm{e}}{1 - C_\mathrm{e}} \frac{m}{a} \right) = 0. \nonumber
\end{eqnarray}
When $\varepsilon_\mathrm{i} = \varepsilon_\mathrm{e} = 0$, one gets the well-known dispersion relation of the MHD normal modes propagating along a flowing cool plasma.

\section{Numerical solutions and discussion}
\label{sec:numerics}
As a model of the moving magnetic flux tube of cool magnetized plasma (see Figure~\ref{fig:fig1}) we use the slow solar wind at a speed of $300$~km\,s$^{-1}$ over streamers, with electron number density $n_\mathrm{i} = 2.5 \times 10^6$~m$^{-3}$, magnetic field $B_0 = 7 \times 10^{-5}$~G, Alfv\'en speed $v_\mathrm{Ai} = 96.5$~km\,s$^{-1}$, and sound speed $c_\mathrm{si} = 40.6$~km\,s$^{-1}$ (at electron temperature $T_\mathrm{e} = 1.2 \times 10^5$~K).  The ion cyclotron frequency is $\omega/2\pi = 106.4$~mHz and accordingly the ion inertial length is $l_\mathrm{Hall} = 144$~km.  In studying the propagation characteristics of Hall MHD waves and their stability, we consider three cases characterized by the density contrast $\eta = \rho_\mathrm{e}/\rho_\mathrm{i}$, equal correspondingly to $0.4$, $0.8$, and $2$.  We obtain these density contrasts by keeping the plasma density inside the tube $n_\mathrm{i}$ constant and varying its value in the environment $n_\mathrm{e}$ from $1.0 \times 10^6$ through $2.0 \times 10^6$ to $5.0 \times 10^6$~m$^{-3}$.  We also assume that the sound speed outside the tube possesses the same value of $40.6$~km\,s$^{-1}$.  In all three cases the pressure balance equation (equality of the sum of thermal and magnetic pressures in both media) yields a ratio of external to internal magnetic field close to $1$, which justifies our assumption to use one background magnetic field $B_0$.  We note also that the plasma betas are less (or much less) than one, thus allowing us to treat both media as cool plasmas.

Prior to beginning the numerical task of solving dispersion equation~(\ref{eq:dispeqn}) in complex variables (assuming a complex angular wave frequency $\omega = \mathrm{Re}\omega + \mathrm{i}\mathrm{Im}\omega$ and a real axial wave number $k_z$) for given mode, say for $m = 1$, that is, the kink mode, it is instructive to explore how the wave dispersion curves and wave growth rates (when the studied mode is unstable) change by considering both media as compressible and cool standard MHD plasmas and cool Hall MHD plasmas, respectively.  To this end, here we give the well-known wave dispersion equation of the normal MHD modes of flowing compressible magnetized plasmas \citep{homem2003,nakariakov2007,zhelyazkov2012}:
\begin{equation}
\label{eq:dispeq}
        \frac{\rho_{\rm e}}{\rho_{\rm i}}\left( \omega^2 - k_z^2 v_{\rm Ae}^2
        \right) m_{0{\rm i}}\frac{I_m^{\prime}(m_{0{\rm
        i}}a)}{I_m(m_{0{\rm i}}a)}
        - \left[ \left( \omega - \vec{k} \cdot
        \vec{v}_0 \right)^2 - k_z^2 v_{\rm Ai}^2 \right] m_{0{\rm
        e}}\frac{K_m^{\prime}(m_{0{\rm e}}a)}{K_m(m_{0{\rm e}}a)} = 0,
\end{equation}
where the squared wave attenuation coefficients in both media are given by the expression
\[
    m_0^2 = -\frac{\left( \Omega^2 - k_z^2 c_{\rm s}^2 \right)\left( \Omega^2 - k_z^2 v_{\rm A}^2 \right)}{\left( c_{\rm s}^2 + v_{\rm A}^2 \right)\left( \Omega^2 - \omega_{\rm c}^2 \right)},
\]
in which $\Omega \equiv \omega$ in the environment, and the cusp frequency, $\omega_\mathrm{c}$, is usually expressed via the 
so-called tube speed, $c_\mathrm{T}$, notably $\omega_\mathrm{c} = k_z  c_\mathrm{T}$, where \citep{edwin1983}
\[
    c_\mathrm{T} = \frac{c_{\rm s}v_{\rm A}}{\sqrt{c_{\rm s}^2 + v_{\rm A}^2}}.
\]
We recall that for the kink mode ($m = 1$) one defines the so-called kink speed \citep{edwin1983},
\begin{equation}
\label{eq:kinkspeed}
        c_{\rm k} = \left( \frac{\rho_{\rm i} v_{\rm Ai}^2 + \rho_{\rm e}
        v_{\rm Ae}^2}{\rho_{\rm i} + \rho_{\rm e}} \right)^{1/2},
\end{equation}
which, as seen, is independent of sound speeds and characterizes the propagation of transverse perturbations.  We will show, that notably the kink mode can become unstable against the KH instability.

In the cold-plasma approximation, when the sound speed $c_\mathrm{s}$ and cusp frequency $\omega_\mathrm{c}$ are equal to zero, the above dispersion equation (\ref{eq:dispeq}) reduces to the form
\begin{equation}
\label{eq:dispeqcold}
        \frac{\rho_{\rm e}}{\rho_{\rm i}}\left( \omega^2 - k_z^2 v_{\rm Ae}^2
        \right) m_\mathrm{0i}^\mathrm{c}\frac{I_m^{\prime}(m_\mathrm{0i}^\mathrm{c}a)}{I_m(m_\mathrm{0i}^\mathrm{c}a)}
        - \left[ \left( \omega - \vec{k} \cdot
        \vec{v}_0 \right)^2 - k_z^2 v_{\rm Ai}^2 \right] m_\mathrm{0e}^\mathrm{c}\frac{K_m^{\prime}(m_\mathrm{0e}^\mathrm{c}a)}{K_m(m_\mathrm{0e}^\mathrm{c}a)} = 0,
\end{equation}
where now the wave attenuation coefficients are given by
\[
    m_\mathrm{0i}^\mathrm{c} = k_z \left[ 1 - (\omega - \vec{k}\cdot \vec{v}_0)^2/k_z^2 v_\mathrm{Ai}^2 \right]^{1/2} \quad \mbox{and} \quad m_\mathrm{0e}^\mathrm{c} = k_z \left( 1 - \omega^2/k_z^2 v_\mathrm{Ae}^2 \right)^{1/2}.
\]

The numerical solving of Equations (\ref{eq:dispeq}), (\ref{eq:dispeqcold}), and (\ref{eq:dispeqn}) will be performed in dimensionless variables.  Thus, we normalize all velocities with respect to the Alfv\'en speed inside the flux tube, $v_\mathrm{Ai}$, and the wavelength $\lambda = 2\pi/k_z$ with respect to the tube radius $a$, which implies that we shall look for solutions of the normalized complex wave phase velocity $\omega/k_z v_\mathrm{Ai}$ as a function of the normalized wavenumber $k_z a$ and input parameters, whose number depends upon the form of the wave dispersion relation.  When we explore Equation~(\ref{eq:dispeq}), along with the density contrast $\eta$, we should evaluate the reduced plasma betas, $\tilde{\beta}_\mathrm{i,e} = c_\mathrm{i,e}^2/v_\mathrm{Ai,e}^2$, magnetic field ratio, $b = B_\mathrm{e}/B_\mathrm{i}$, and Alfv\'en Mach number $M_\mathrm{A} = v_0/v_\mathrm{Ai}$, which represents the flow velocity ${v}_0$.  We note that for normalization of the sound speeds one needs the parameters $\tilde{\beta}_\mathrm{i,e}$, while $b$ is used at  the normalization of the Alfv\'en speed in the environment, $v_\mathrm{Ae}$.  For cold plasmas we take $b = 1$.

At given sound and Alfv\'en speeds, alongside the input parameters $\eta$ and $b$, one can make some predictions, namely to specify the nature of the mode (pure surface, pseudosurface/body, or leaky) \citep{cally1986}, the value of the kink speed (\ref{eq:kinkspeed}) in a static flux tube, and the expected threshold/critical Alfv\'en Mach number at which KHI would start---the latter is determined by the inequality \citep{zaqarashvili2014}
\begin{equation}
\label{eq:criterion}
    |m|M_\mathrm{A}^2 > (1 + 1/\eta)(|m|b^2 + 1).
\end{equation}

In the next subsection we present the results of the numerical task for the dispersion and growth rate curves of the kink ($m = 1$) mode in moving magnetic flux tubes with different values of the density contrast $\eta$.

\subsection{Kink mode propagation characteristics at density contrasts $0.4$, $0.8$, and $2$}
\label{subsec:kink}

At $\eta = 0.4$, the Alfv\'en speed outside the flux tube (computed from the pressure balance equation) is $v_\mathrm{Ae} = 162$~km\,s$^{-1}$, the magnetic field ratio $b = 1.062$, and plasma betas are $\beta_\mathrm{i} = 0.213$ and $\beta_\mathrm{e} = 0.076$, respectively.  The ordering of sound and Alfv\'en speeds in the system is as follows:
\[
    c_\mathrm{i} = c_\mathrm{e} < v_\mathrm{Ai} < v_\mathrm{Ae}.
\]
According to \citet{cally1986} (see Table I there), in such a case the propagating kink mode must be a pseudosurface/body wave of B$^{+}_{+}$ type.  The kink speed, calculated from (\ref{eq:kinkspeed}), is equal to ${\approx}119$~km\,s$^{-1}$; thus its normalized value is $c_\mathrm{k}/v_\mathrm{Ai} = 1.2327$.  The threshold Alfv\'en Mach number at which the KHI should occur is equal to $2.729$.  With input parameters: $\eta = 0.4$, $\tilde{\beta}_\mathrm{i} = 0.1773$, $\tilde{\beta}_\mathrm{e} = 0.0629$,
$b = 1.062$, and various magnitudes of the Alfv\'en Mach number $M_\mathrm{A}$ we obtain a
\begin{figure}[!h]
   \centerline{\hspace*{0.015\textwidth}
               \includegraphics[width=0.515\textwidth,clip=]{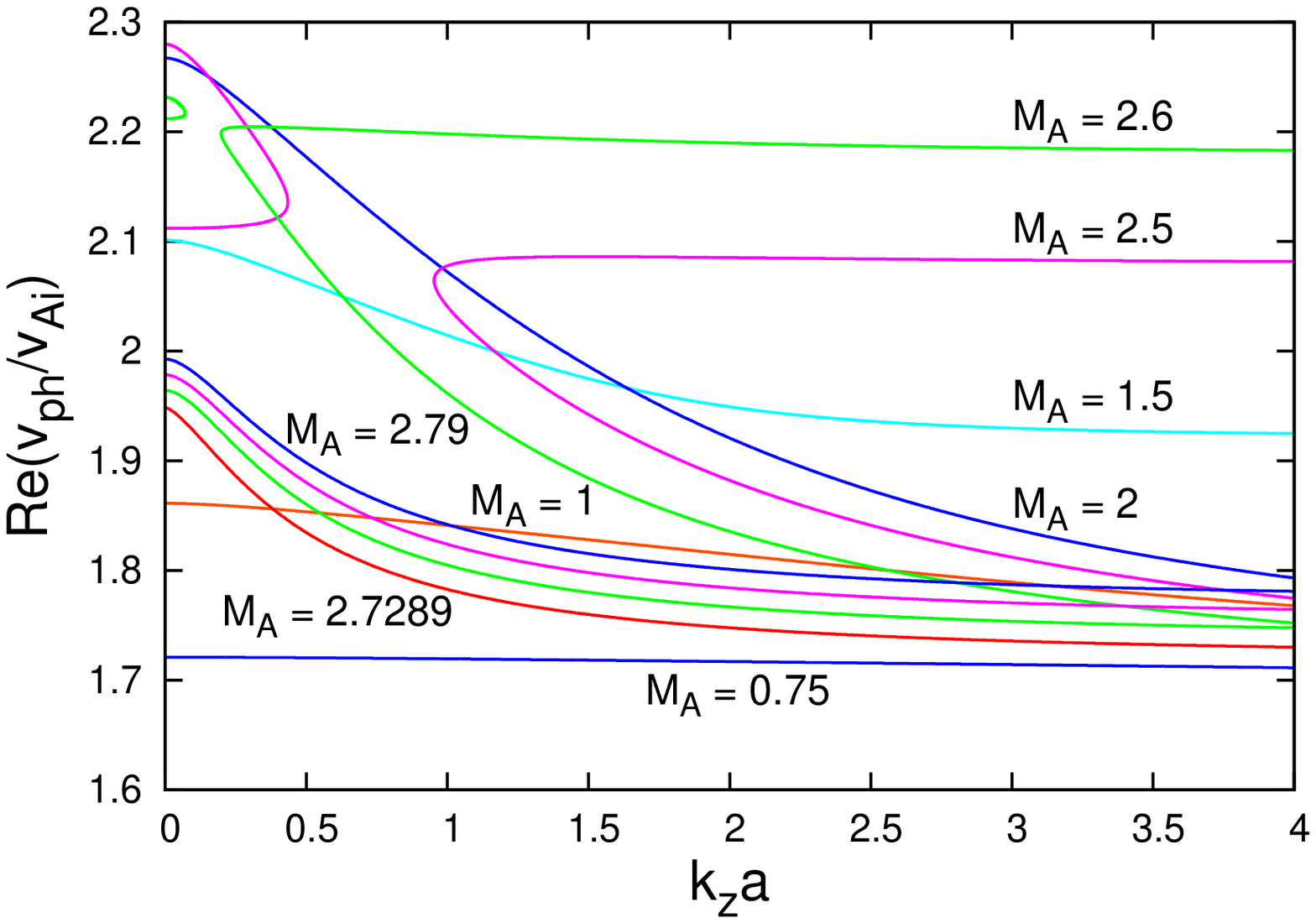}
               \hspace*{-0.03\textwidth}
               \includegraphics[width=0.515\textwidth,clip=]{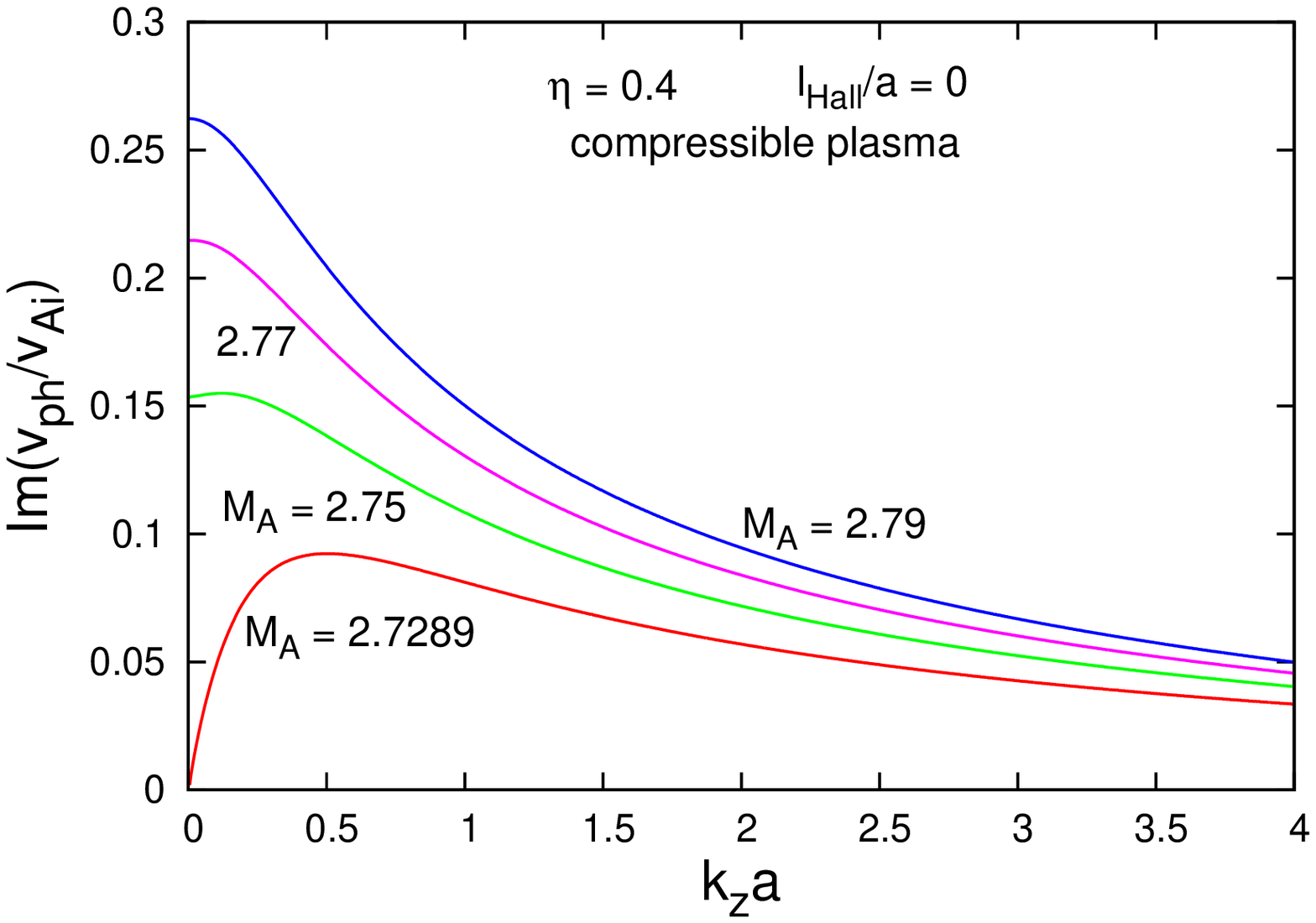}
              }
  \caption{(\emph{Left panel}) Dispersion curves of stable and unstable kink ($m = 1$) MHD mode propagating in a moving magnetic flux tube of compressible plasma at $\eta = 0.4$ and $b = 1.06$.  Unstable dispersion curves, located above the $M_\mathrm{A} = 0.75$-dispersion curve, have been calculated for four values of the Alfv\'en Mach number $M_\mathrm{A} = 2.7289$, $2.75$, $2.77$, and $2.79$.  (\emph{Right panel}) The normalized growth rates of the unstable mode for the same values of $M_\mathrm{A}$.  \emph{Red\/} curves in both plots correspond to the onset of KH instability.}
   \label{fig:fig2}
\end{figure}
\begin{figure}[!ht]
   \centerline{\hspace*{0.015\textwidth}
               \includegraphics[width=0.515\textwidth,clip=]{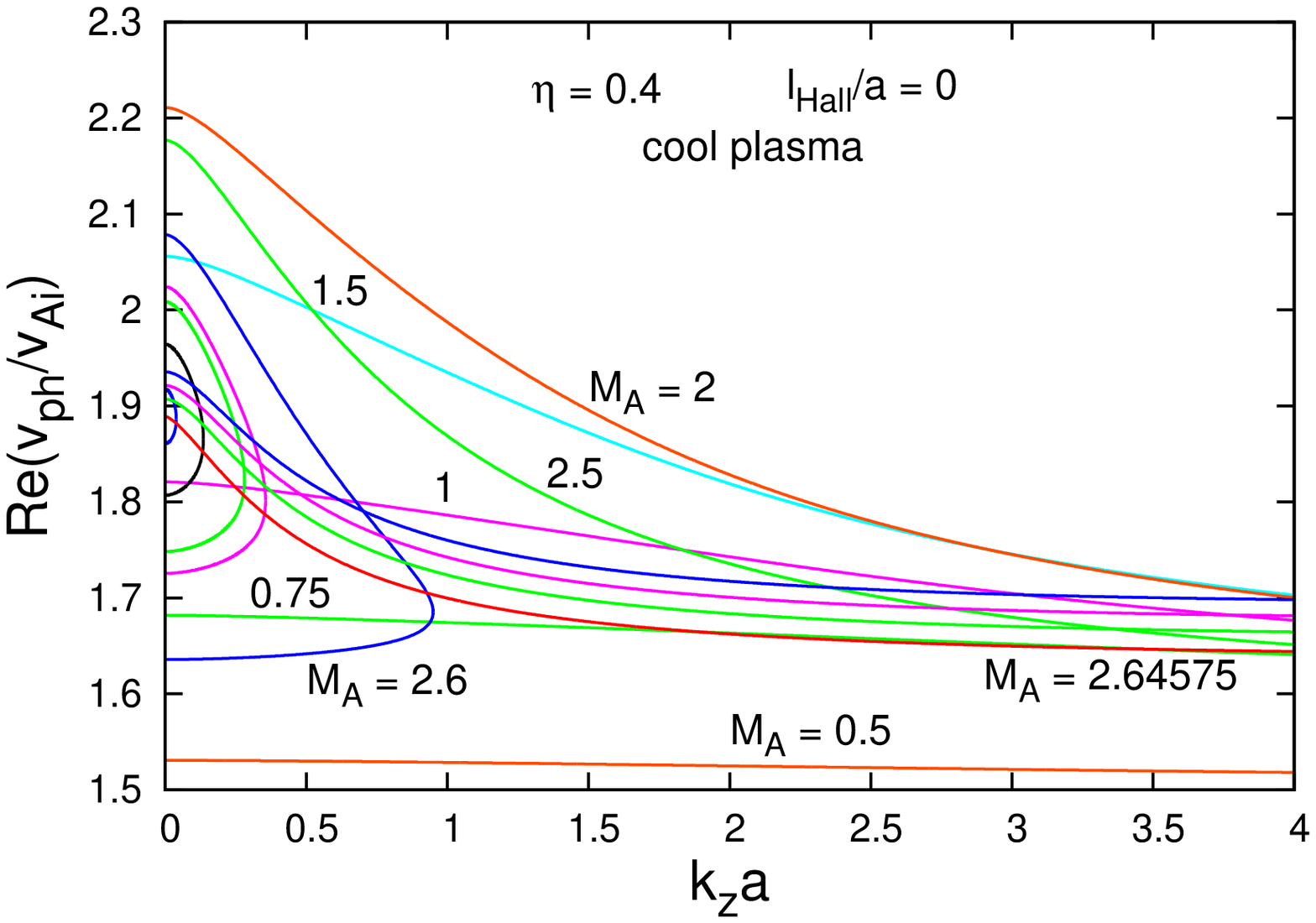}
               \hspace*{-0.03\textwidth}
               \includegraphics[width=0.515\textwidth,clip=]{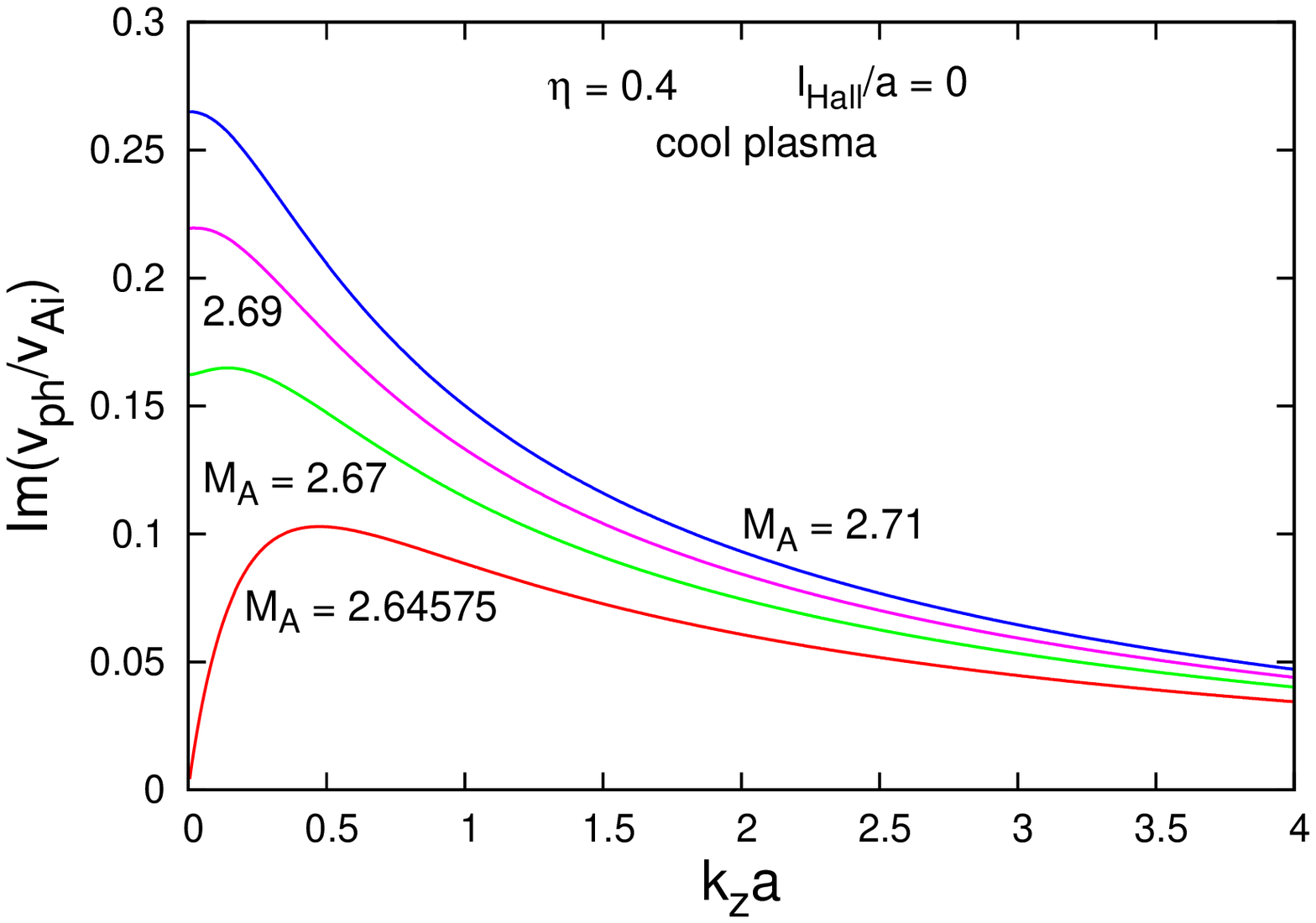}
              }
  \caption{The same as in Figure~\ref{fig:fig2}, but with $b = 1$.  Unstable dispersion curves, located above the $M_\mathrm{A} = 0.5$-dispersion curve, have been calculated for $M_\mathrm{A} = 2.64575$, $2.67$, $2.69$, and $2.71$.}
   \label{fig:fig3}
\end{figure}
set of dispersion curves (of both stable and unstable waves) and normalized growth rate
curves shown in Figure~\ref{fig:fig2}.  We would like
to note that the normalized kink-speed dispersion curve (not shown in Figure~\ref{fig:fig2}) appears exactly at the predicted value of $1.2327$.  This kink mode is really a pseudosurface wave in agreement with Cally's criterion.
With including the flow, that curve splits into two curves \citep{zhelyazkov2012} which at small Alfv\'en Mach numbers $M_\mathrm{A}$ go almost parallel, but for larger values of $M_\mathrm{A}$ there shapes dramatically change.  This is especially true in the region of normalized wave velocities where one expects a KHI onset (see, for instance, the curves for $M_\mathrm{A} = 2.5$ and $2.6$)---according to the criterion (\ref{eq:criterion}) that should happen at $M_\mathrm{A} > 2.729$.  Our computations show that the marginally dispersion and growth rate curves appear at $M_\mathrm{A} = 2.7289$ which is in excellent agreement with the predicted value.

It is curious to see what patterns of dispersion and growth rate curves one will obtain when using Equation~(\ref{eq:dispeqcold}) describing the wave propagation in a cool moving magnetic flux tube.  With two input parameters, namely $\eta = 0.4$ and $b = 1$,
one obtains the curves shown in Figure~\ref{fig:fig3}.  As seen, the patterns are similar (in a sense) with those shown in
Figure~\ref{fig:fig2}.  Starting at $M_\mathrm{A} = 2.6$ the pair of kink-speed dispersion curves merge forming
a semi-closed curve, which becomes narrower with the increase in $M_\mathrm{A}$: the non-labeled purple curve is calculated at $M_\mathrm{A} = 2.625$, the green one at $2.63$, the black semi-closed curve at $2.64$, and the blue one at $M_\mathrm{A} = 2.645$.  KHI arises at the threshold Alfv\'en Mach number $M_\mathrm{A} = 2.64575$, which is lower than the predicted one, but generally on the same order.  In both cases (of compressible and cool plasma) the critical flow velocity
at which the instability starts is ${\cong}263$ or $255$~km\,s$^{-1}$, being lower than the assumed flow
speed of $300$~km\,s$^{-1}$.  In other words, in both approximations the moving cool-plasma magnetic flux tube is unstable against the KHI.  The similarity of wave dispersion and growth rate curves patterns shown in Figures~\ref{fig:fig2} and \ref{fig:fig3} gives us reason to use the cool-plasma approximation as a reliable one.  The next logical step is to see how the Hall MHD will change the picture.
\begin{figure}[!ht]
   \centerline{\hspace*{0.015\textwidth}
               \includegraphics[width=0.515\textwidth,clip=]{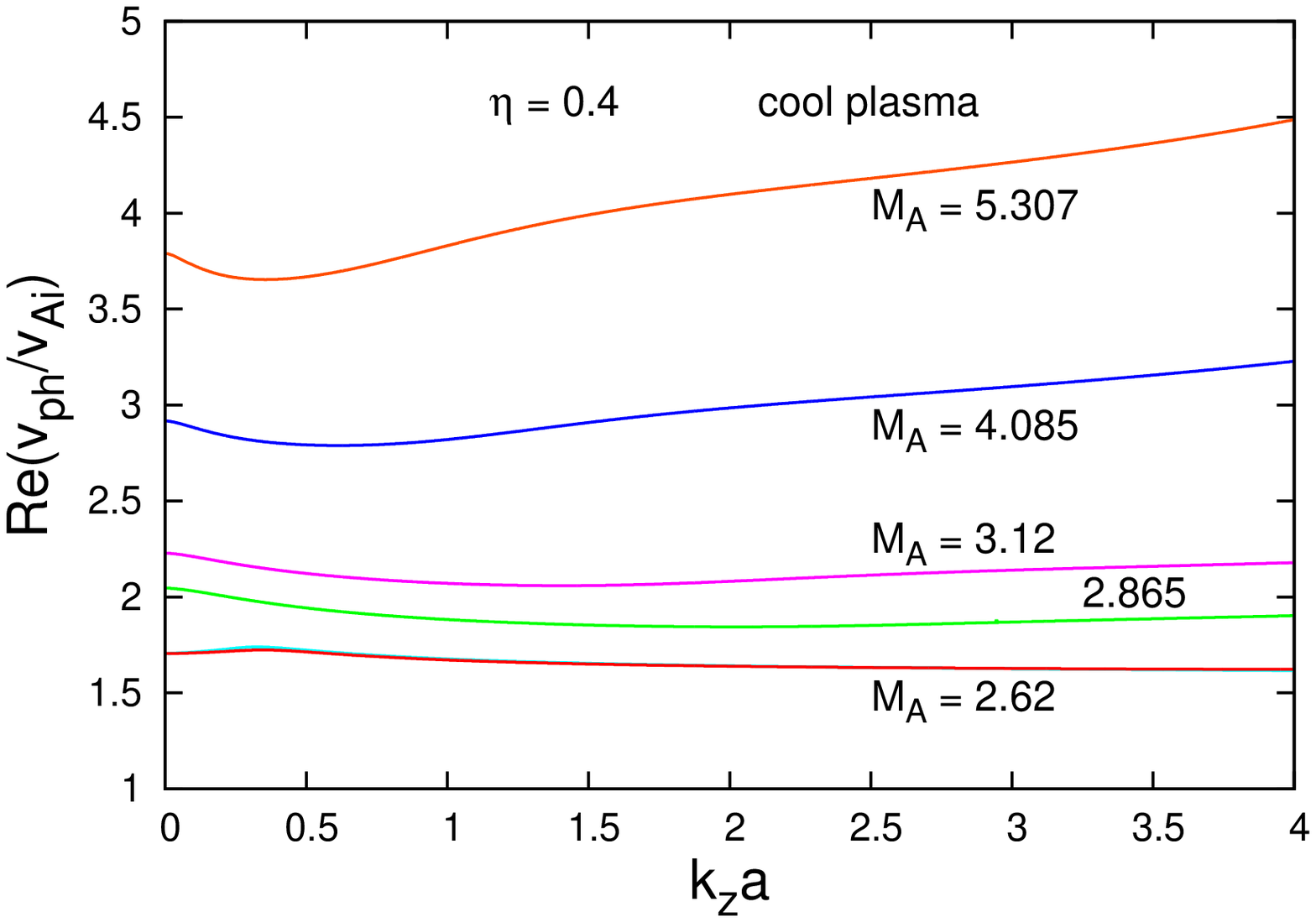}
               \hspace*{-0.03\textwidth}
               \includegraphics[width=0.515\textwidth,clip=]{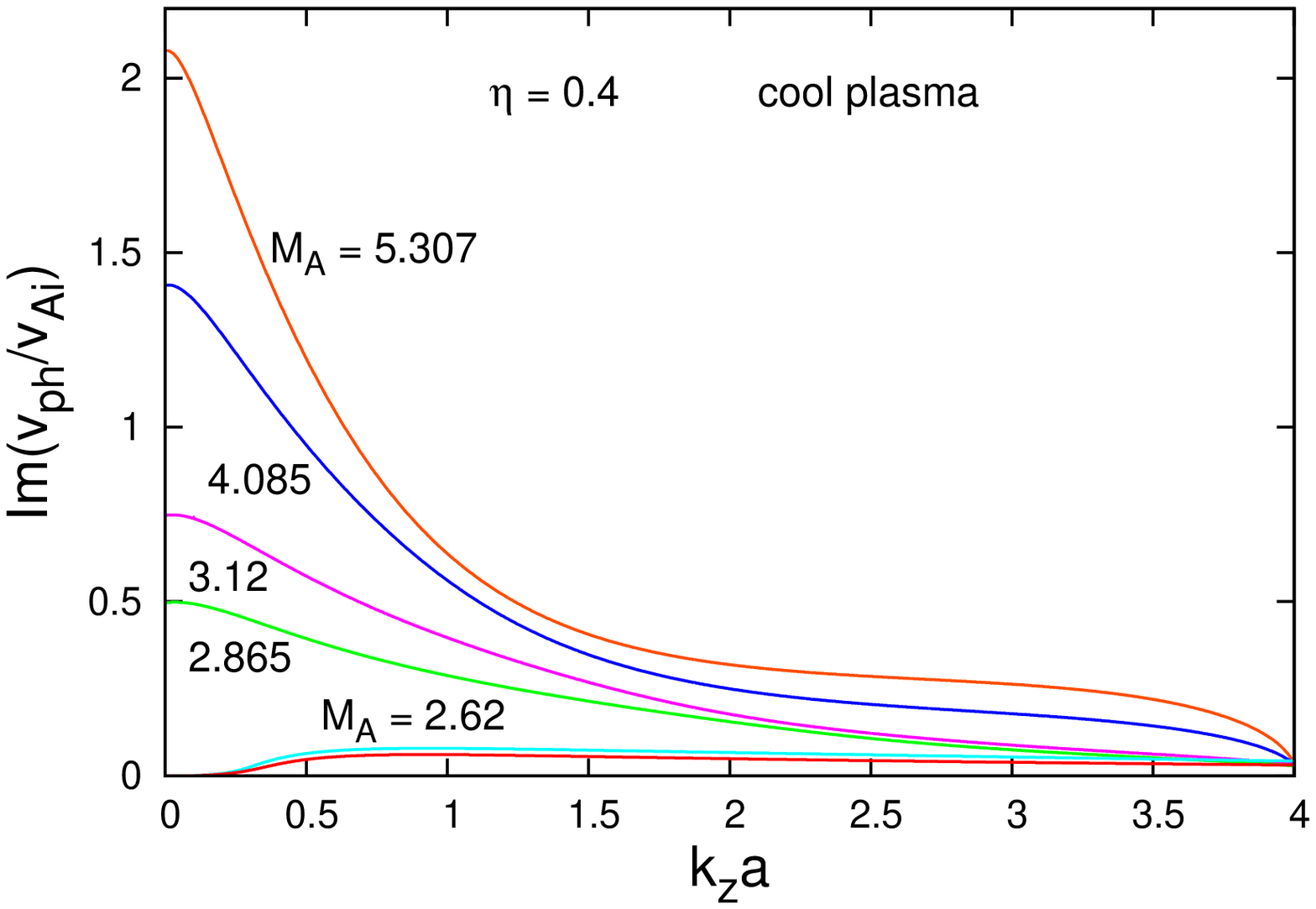}
              }
  \caption{(\emph{Left panel}) Dispersion curves of unstable kink ($m = 1$) Hall MHD mode propagating in a moving magnetic flux tube of cool plasma at $\eta = 0.4$, $b = 1$, and at various values of the parameter $l_\mathrm{Hall}/a$, notably equal to $0.0025$, $0.01$, $0.05$, $0.1$, $0.25$, and $0.4$. Threshold Alfv\'en Mach numbers for the corresponding values of $l_\mathrm{Hall}/a$ are as follows: $2.62$, $2.62$, $2.865$, $3.12$, $4.085$, and $5.307$.  (\emph{Right panel}) The normalized growth rates of the unstable Hall MHD mode for the same values of the input parameters.}
   \label{fig:fig4}
\end{figure}

The normalization of the parameters $\varepsilon_\mathrm{i}$ and $\varepsilon_\mathrm{e}$ in Equation~(\ref{eq:dispeqn}) requires the usage of the numerical parameter $l_\mathrm{Hall}/a$.  Thus, in finding the solutions to Equation~(\ref{eq:dispeqn}) in addition to input parameters $\eta$, $b$, and $M_\mathrm{A}$, we have to specify the value of that ratio $l_\mathrm{Hall}/a$.  The algorithm in solving the Hall MHD dispersion relation is the following: for a fixed value of $l_\mathrm{Hall}/a$ we have to find that threshold Alfv\'en Mach number at which KHI rises, that is, to obtain the marginal dispersion and growth rate curves (the red curves in Figures~\ref{fig:fig2} and \ref{fig:fig3}).  Our choice for the set of $l_\mathrm{Hall}/a$-values is: $0.0025$, $0.01$, $0.05$, $0.1$, $0.25$, and $0.4$.  The results of the numerical calculations are presented in Figure~\ref{fig:fig4}.  It is easily seen that the small $l_\mathrm{Hall}/a$-values do not change notably both the dispersion and the growth rate curves compared with those obtained in the frame of the standard MHD (look at Figure~\ref{fig:fig3}).  For bigger values of the same parameter, one observes distinct increases in the normalized wave phase velocity and the corresponding dimensionless growth rate.  Furthermore, at $l_\mathrm{Hall}/a = 0.1$ the KHI occurs at a threshold Alfv\'en Mach number equal to $3.12$.  This means that the critical flow velocity for the instability onset is $v_0^\mathrm{cr} \cong 301$~km\,s$^{-1}$.  If we assume that the tube radius is $a = 1000$~km, then with $l_\mathrm{Hall} = 114$~km the magnitude of $l_\mathrm{Hall}/a = 0.114$ is sufficient to suppress the KHI.  In other words, under these circumstances the Hall term stabilizes the plasma flow.

With the increased value of the parameter $\eta = 0.8$, for compressible plasma the pressure balance equation yields $v_\mathrm{Ae} \cong 110$~km\,s$^{-1}$ and $b = 1.021$.  In that case the reduced plasma betas are
$\tilde{\beta}_\mathrm{i} = 0.1773$ and $\tilde{\beta}_\mathrm{e} = 0.1361$, respectively, while the normalized kink speed is equal to $1.0652$, that is, $c_\mathrm{k} \cong 103$~km\,s$^{-1}$.  The ordering of sound and Alfv\'en speeds is the same as in the previous case of $\eta = 0.4$, which implies that the kink mode should be a pseudosurface/body wave of B$^{+}_{+}$ type \citep{cally1986}.  The threshold Alfv\'en Mach number for instability onset, according to inequality (\ref{eq:criterion}), has to be higher than $2.1437$.  The numerical solutions to the dispersion equation~(\ref{eq:dispeq}) yield dispersion and growth rate curve patterns similar to those shown in Figure~\ref{fig:fig2}.  The numerical code reproduces the normalized kink speed to its expected value within four places behind the decimal point, that is $c_\mathrm{k}/v_\mathrm{Ai} = 1.0652$.  The numerically found threshold Alfv\'en Mach number equals $2.1437$ which yields a critical flow velocity of ${\cong}207$~km\,s$^{-1}$, which means that in this standard MHD approximation of compressible plasma the kink mode ($m = 1$) is definitely unstable against KHI.  In the same standard MHD, but in the cold approximation, the solutions to Equation~(\ref{eq:dispeqcold}) yield an even lower threshold $M_\mathrm{A} = 2.12$, that is, a flow speed of $204.6$~km\,s$^{-1}$ is sufficient for the triggering of KHI in the system.  However, the big surprise comes with the solutions to the Hall MHD dispersion %
\begin{figure}[!ht]
   \centerline{\hspace*{0.015\textwidth}
               \includegraphics[width=0.515\textwidth,clip=]{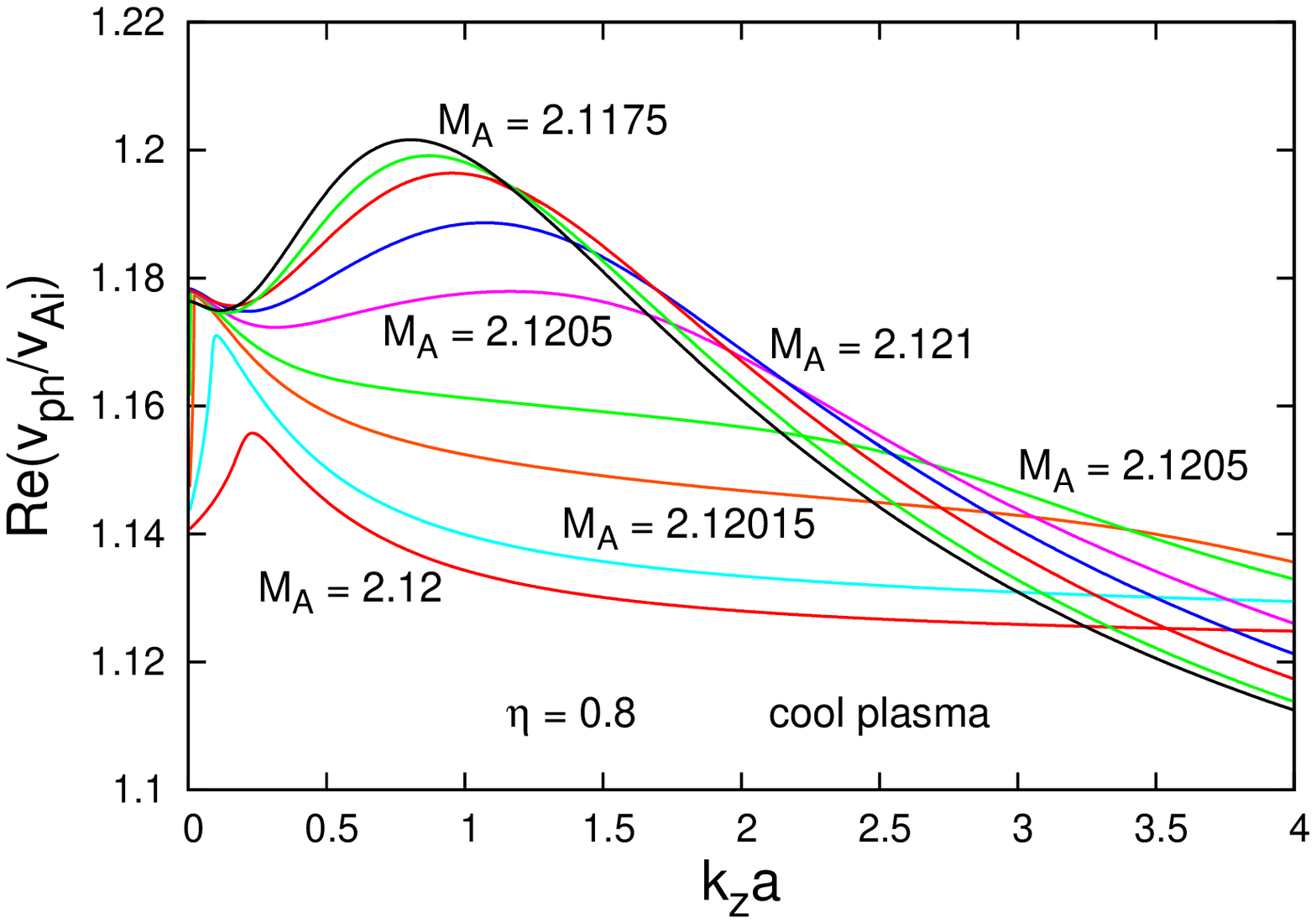}
               \hspace*{-0.03\textwidth}
               \includegraphics[width=0.515\textwidth,clip=]{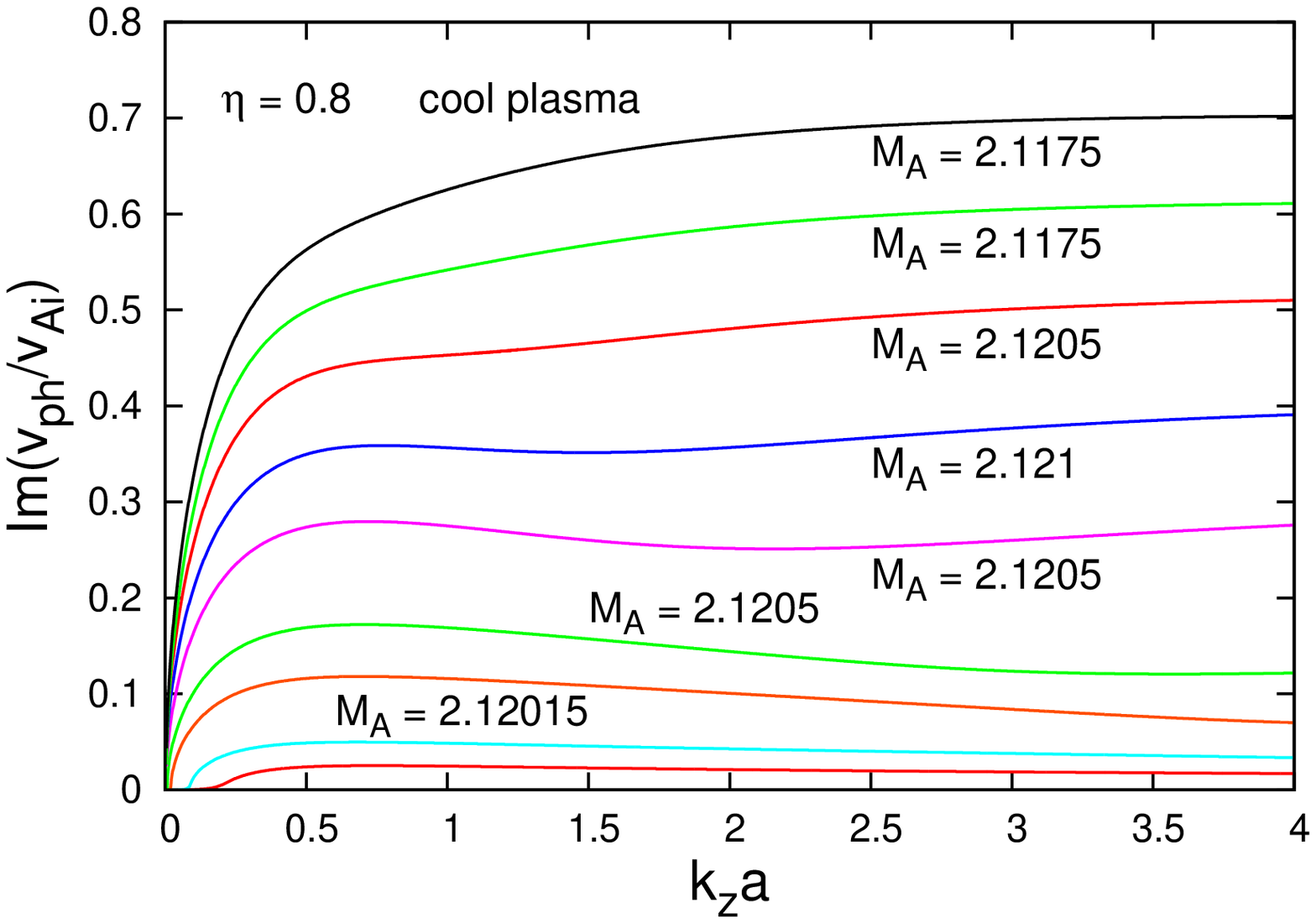}
              }
  \caption{The same as in Figure~\ref{fig:fig4}, but for $\eta = 0.8$ and at $l_\mathrm{Hall}/a$ equal to $0.0025$, $0.01$, $0.05$, $0.1$, $0.25$, $0.4$, $0.6$, $0.8$, and $1$.  Threshold Alfv\'en Mach numbers for the corresponding values of $l_\mathrm{Hall}/a$ are as follows: $2.12$, $2.12015$, $2.12015$, $2.1205$, $2.1205$, $2.121$, $2.1205$, $2.1175$ and $2.1175$.}
   \label{fig:fig5}
\end{figure}
Equation~(\ref{eq:dispeqn})---the plots of marginal dispersion and growth rate curves of the unstable kink mode for an extended set of values of the parameter $l_\mathrm{Hall}/a$ are presented in Figure~\ref{fig:fig5}.
It is immediately seen how different are the curves' patterns---compare Figures~\ref{fig:fig4} and \ref{fig:fig5}---and what is really unexpected, any increase in the $l_\mathrm{Hall}/a$-value does not increase the threshold Alfv\'en Mach number for the instability occurrence---just the opposite: the large enough $l_\mathrm{Hall}/a$-values of $0.8$ and $1$ yield lower threshold $M_\mathrm{A} = 2.1175$ than that obtained for $l_\mathrm{Hall}/a = 0$ or $0.0025$.  With $M_\mathrm{A} = 2.1175$, the critical flow velocity for KHI onset is ${\cong}204$~km\,s$^{-1}$, which is roughly $100$~km\,s$^{-1}$ lower than the assumed maximal slow-solar-wind speed of $300$~km\,s$^{-1}$.

\begin{figure}[!ht]
   \centerline{\hspace*{0.015\textwidth}
               \includegraphics[width=0.515\textwidth,clip=]{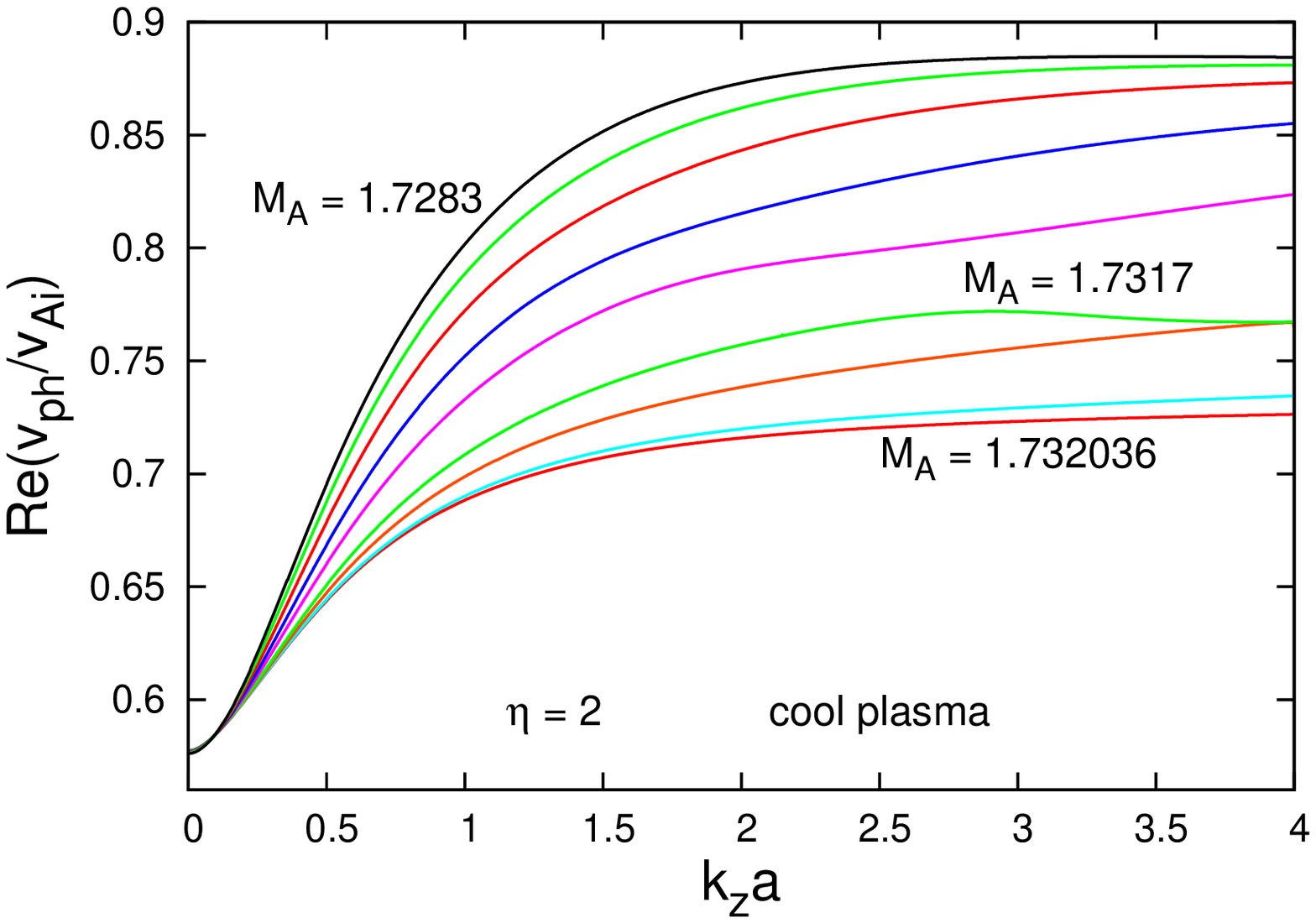}
               \hspace*{-0.03\textwidth}
               \includegraphics[width=0.515\textwidth,clip=]{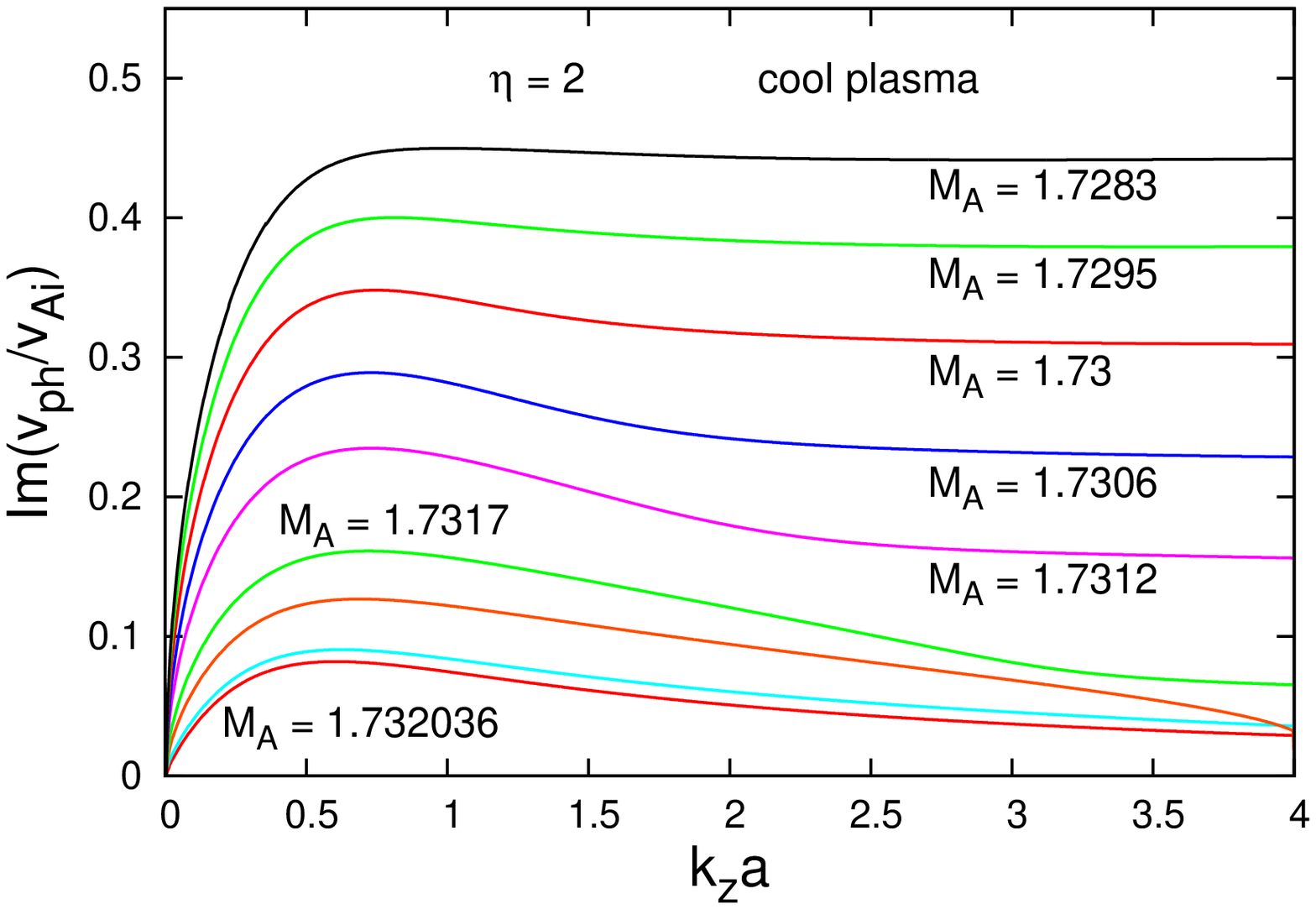}
              }
  \caption{The same as in Figure~\ref{fig:fig5} at $\eta = 2$ and the same values of $l_\mathrm{Hall}/a$.  Threshold Alfv\'en Mach numbers for the corresponding values of $l_\mathrm{Hall}/a$ are as follows: $1.732036$, $1.732037$, $1.7318$, $1.7317$, $1.7312$, $1.7306$, $1.73$, $1.7295$ and $1.7283$.}
   \label{fig:fig6}
\end{figure}

A further increase in the parameter $\eta$ by choosing it equal to $2$, yields a completely different picture.  First and foremost, in the compressible plasma approximation, the pressure balance equation gives a rather low value of $v_\mathrm{Ae} = 60.5$~km\,s$^{-1}$ and consequently a $b = 0.8872$, which implies the propagation of sub-Alfv\'enic kink waves---the normalized kink speed is equal to $0.77183$, which yields $c_\mathrm{k} = 74.5$~km\,s$^{-1}$.  Moreover, now the ordering of sound and Alfv\'en speeds of the system has changed and that chain has the form
\[
        c_\mathrm{i} = c_\mathrm{e} < v_\mathrm{Ae} < v_\mathrm{Ai}.
\]
Such an ordering does not fit any of possible non-leaky modes listed in Table I in \citealp{cally1986}.  With
$\tilde{\beta}_\mathrm{i} = 0.1773$ and $\tilde{\beta}_\mathrm{e} = 0.4506$, at $M_\mathrm{A} = 0$ (static plasma) the numerical solution to Equation~(\ref{eq:dispeq}) reproduces the normalized kink speed up to four places behind the decimal point, but what
is more important, one finds that in this case the kink ($m = 1$) mode is a leaky wave (real $m_\mathrm{0i}$ and imaginary $m_\mathrm{0e}$).  The dispersion curves' pattern is much more complicated than that of the previous two cases and the threshold Alfv\'en Mach number of $1.637314$ found is very close to the predicted one (${=}1.637$).  Hence, a flow speed of $158$~km\,s$^{-1}$ ensures the triggering of the KHI in the system.  In the cold-plasma approximation, that threshold $M_\mathrm{A}$-value is a little bit higher (${=}1.732045$), but still giving a relatively low flow speed of $167$~km\,s$^{-1}$ for the instability onset.  The Hall MHD approach for the same set of values for the $l_\mathrm{Hall}/a$ parameter as in the previous case of $\eta = 0.8$ yields more or less similar pattern of the growth rate curves (see the right panel in Figure~\ref{fig:fig6}), but different for
wave dispersion curves.  The latter have two specific issues: (i) all dispersion curves begin with a normalized phase velocity of ${\approx}0.577$, or equivalently with ${\cong}55.7$~km\,s$^{-1}$; (ii) all dispersion curves, even those corresponding to relatively big $l_\mathrm{Hall}/a$-values like the kink-speed curve in a static compressible standard MHD magnetic flux tube describe sub-Alfv\'enic waves.  Concerning the threshold $M_\mathrm{A}$-value at a small $l_\mathrm{Hall}/a$-value (${=}0.0025$), it is rather close to that in a moving cold-plasma magnetic flux tube---it is equal to $1.732036$, which yields the same critical flow velocity of $167$~km\,s$^{-1}$.  Likewise, in the previous case of $\eta = 0.8$, the kink ($m = 1$) mode is unstable regardless of the value of the $l_\mathrm{Hall}/a$ parameter.

\subsection{Sausage mode propagation characteristics at a density contrast $0.4$}
\label{subsec:sausage}
\begin{figure}[!ht]
   \centerline{\hspace*{0.015\textwidth}
               \includegraphics[width=0.515\textwidth,clip=]{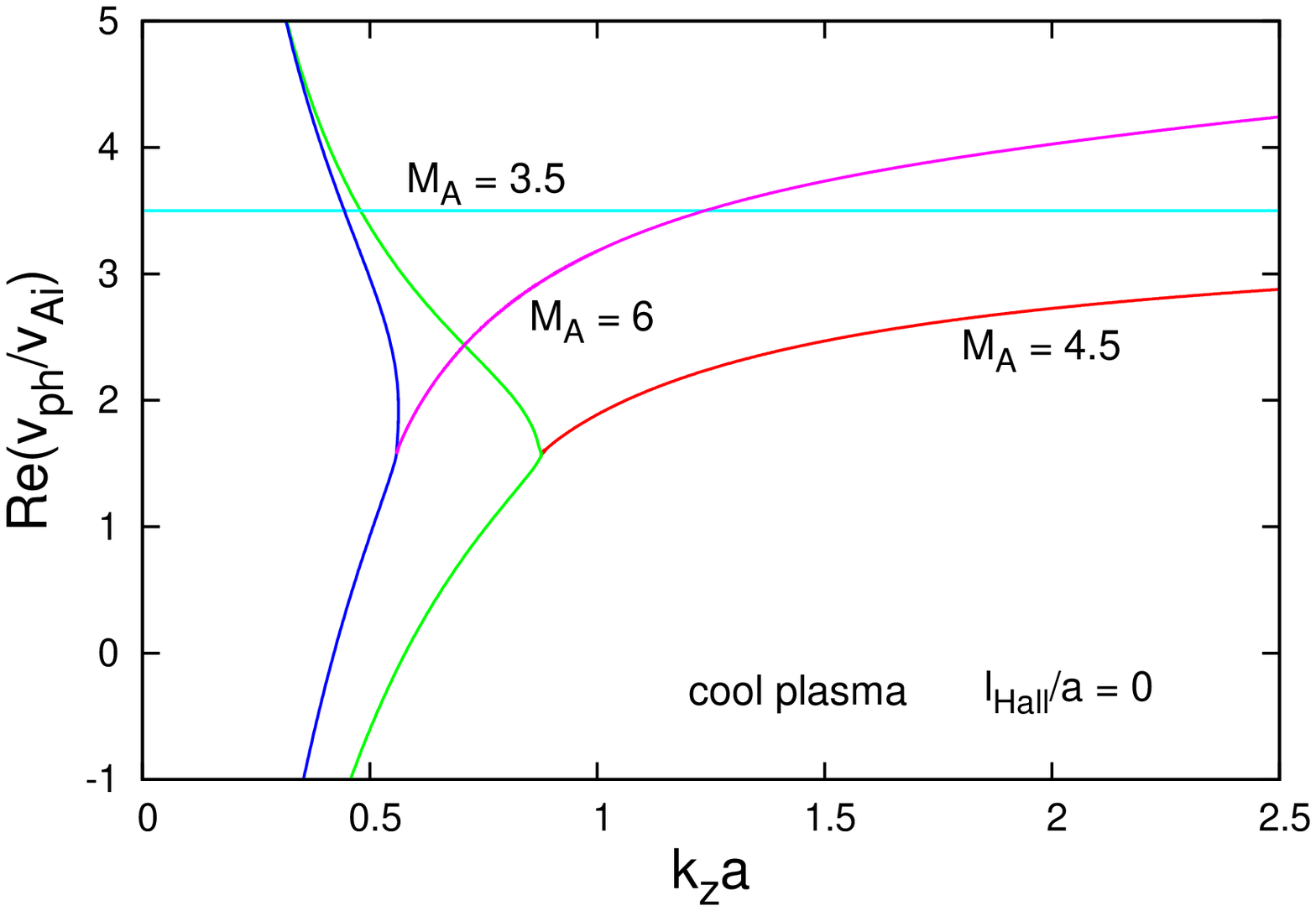}
               \hspace*{-0.03\textwidth}
               \includegraphics[width=0.515\textwidth,clip=]{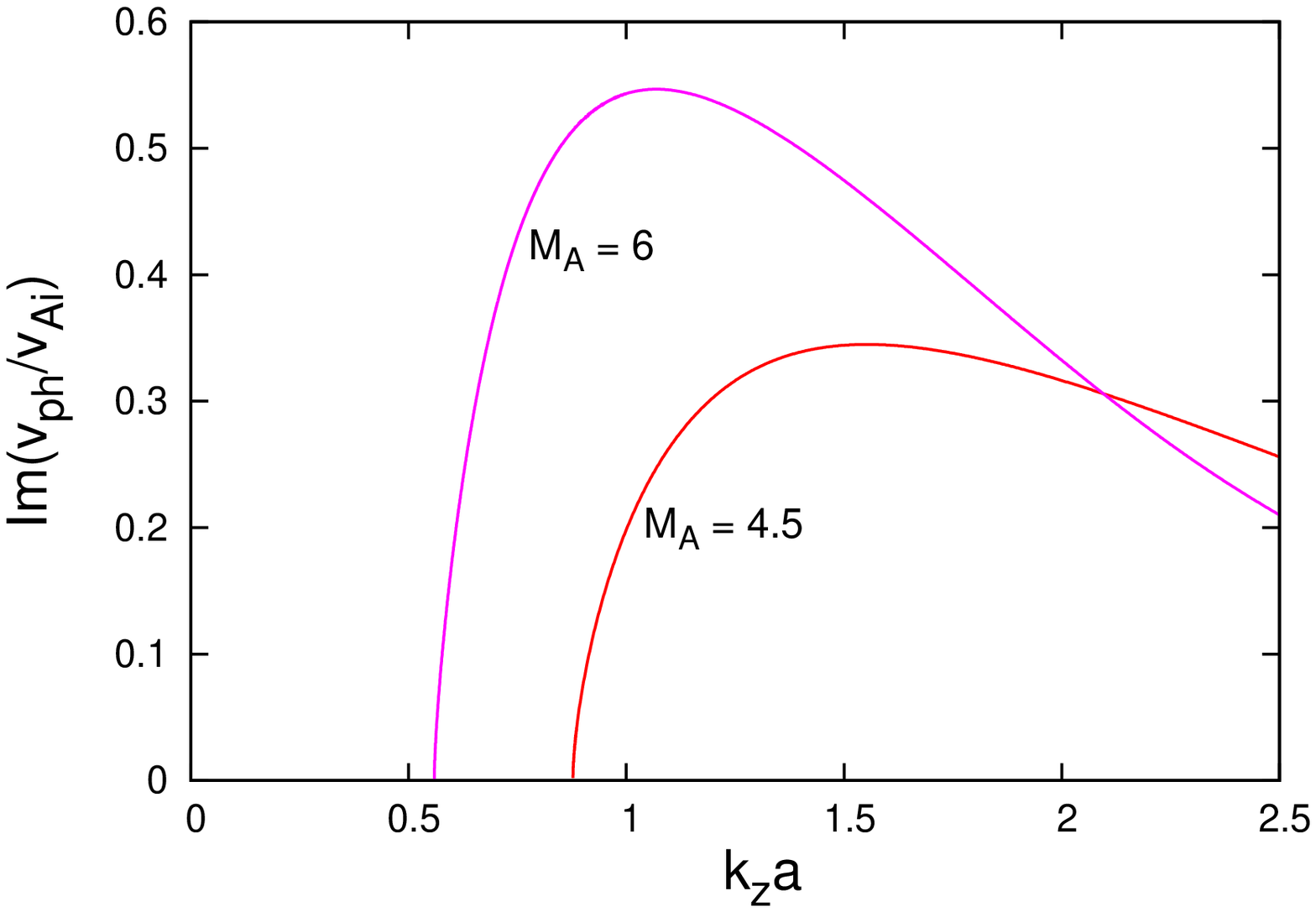}
              }
  \caption{(\emph{Left panel}) Dispersion curves of stable and spuriously unstable sausage ($m = 0$) MHD mode propagating in a moving magnetic flux tube of cool plasma at $\eta = 0.4$, $b = 1$, and at two values of the Alfv\'en Mach numbers equal to $4.5$ and $6$, respectively.  The \emph{green, blue, and cyan curves\/} describe stable wave propagation, while the \emph{green and purple/blueviolet curves\/} simulate spurious unstable sausage waves.  (\emph{Right panel}) The normalized growth rates of the spuriously unstable sausage MHD mode for the same values of the Alfv\'en Mach number.}
   \label{fig:fig7}
\end{figure}

The previous study of \citet{zhelyazkov2010} on Hall MHD modes in the solar wind explored in incompressible plasma approximations has established that in the standard MHD the sausage mode ($m = 0$) is always stable with respect to KHI.  That conclusion has been drawn from the exact solutions to the corresponding dispersion equation, which has the form of a quadratic algebraic equation.  The Hall term in the generalized Ohm law does not change that conclusion either.  We ask ourselves whether the cool-plasma approximation will change
the situation?  The answer to this question is negative.  That can be seen in Figure~\ref{fig:fig7} which presents the results of solving Equation~(\ref{eq:dispeqcold}) at $\eta = 0.4$ and $b = 1$ for two values of the Alfv\'en Mach number equal to $4.5$ and $6$,
respectively.  One obtains two types of dispersion curves, notably curves describing a stable wave propagation (the green and blue curves in Figure~\ref{fig:fig7}) and two other curves (the red and purple ones) which pass through/cross the green and blue ones.  The two latter curves (red and purple) are, however, plotted from spurious real roots of the wave dispersion relation---they do not correspond to real unstable waves.  The same applies for the spurious growth rate curves seen in the right panel of Figure~\ref{fig:fig7}.  These circumstances make us very cautious in the interpretation of numerical solutions of dispersion equations in complex variables.  In fact, the only real dispersion curves are those which are associated with the Alfv\'en speed at given Alfv\'en Mach number.  For example, the cyan curve in the left panel of Figure~\ref{fig:fig7} is the lower of the pair of Alfv\'en-speed curves (at normalized phase velocities of $3.5$ and $5.5$) which are akin to the kink-speed curves discussed in Subsection~\ref{subsec:kink} for $\eta = 0.4$.  It turns out that the inclusion of the Hall term in the Faraday equation~(\ref{eq:faraday}) does not change the above conclusion.

\section{Conclusion}
\label{sec:conclusion}
In this paper, we have studied the propagation of normal modes on a moving magnetic flux tube in three different approximations
(compressible and cold plasma in the standard MHD, and cool plasma in the Hall MHD) for three cases, corresponding to different values of the density contrast $\eta = \rho_\mathrm{e}/\rho_\mathrm{i}$, equal to $0.4$, $0.8$, and $2$, respectively.  The exploration of the wave dispersion equation for a flux tube of compressible plasma allows us to determine the nature of the propagating mode (be it pure surface, pseudosurface/body, or leaky wave) as well as to find the typical speed of the kink ($m = 1$) or sausage ($m = 0$) mode. In this approximation, one can also find the threshold Alfv\'en Mach number $M_\mathrm{A}$ at which the kink ($m = 1$) mode become unstable against the KHI.  The solutions found in the simpler approximation of a cold plasma yield dispersion and growth rate curves' patterns very similar to those obtained for compressible plasmas.  This similarity allows us to use that limit of cool plasma in studying the normal mode propagation in the framework of the Hall MHD.  The main findings in that direction can be summarized as follows:

The Hall term does not change the nature of stable studied modes---depending on the ordering of the sound and Alfv\'en speeds they can be pseudosurface/body or leaky waves.  All unstable waves possess complex attenuation coefficients which circumstance implies that they are neither pseudosurface nor leaky waves---such unstable modes can be classified as generalized surface waves.

The influence of the Hall term on the development of the KHI in spatially bounded flowing structures can be characterized by the parameter $l_\mathrm{Hall}/a$, where $a$ is a scale parameter, equal to the tube radius in the cases of moving cylindrical flux tubes.  For a relatively high density contrast (for instance, at $\eta = 0.4$) the kink ($m = 1$) mode at small $l_\mathrm{Hall}/a$-values can be unstable against the KHI, but a large enough $l_\mathrm{Hall}/a$-value can suppress the instability.  As an illustration, we showed that this can happen for a moving cool-plasma magnetic flux tube with Alfv\'en speed of $96.5$~km\,s$^{-1}$, which is KH unstable for all $l_\mathrm{Hall}/a < 0.1$ but becomes stable for $l_\mathrm{Hall}/a$-values larger than $0.1$.

This situation dramatically changes when the density contrast becomes lower, say, $\eta = 0.8$ or $2$.  In both cases the Hall term does not influence the development of KHI irrespective of the $l_\mathrm{Hall}/a$-value.  Moreover, a high enough $l_\mathrm{Hall}/a$ may diminish the magnitude of the threshold Alfv\'en Mach number $M_\mathrm{A}$ and consequently the corresponding critical flow velocity.

While for $\eta = 0.4$ or $0.8$ all kink waves, stable or unstable, are super-Alfv\'enic waves, at $\eta = 2$ they become sub-Alfv\'enic waves, whose instability status as we already said, is not influenced noticeably by the Hall term.  Even it can facilitate the KHI onset.

The sausage ($m = 0$) mode is not subject to the KHI in all approximations discussed above.

No doubt that a more realistic model of the solar wind would be a $\beta \sim 1$ magnetized flowing plasma.  In that case, the medium can be treated as a nearly incompressible fluid \citep{zank1993}---thus, the building of a correct model in that approximation with taking into account the role of the Hall term will help us to better understand the complex wave phenomena
in the solar wind.  The wave dispersion relation derived for flowing incompressible Hall MHD plasmas should contains the new feature of our finding that in cylindrical geometry one should expect the excitation of MHD modes with arbitrary azimuthal mode number $m$ and it is curious to see which mode is the most unstable against the KHI.  The Kelvin--Helmhotz instability is an important instability because it can trigger the wave turbulence which along with the micro/nano magnetic reconnection, yielding microflares/nanoflares, is considered as one of the main heating mechanisms of the solar corona.

%
\begin{acks}
Our work was supported by the Bulgarian Science Fund under the Indo--Bulgarian project DNTS/INDIA 01/7.  We thank Dr.~Snezhana Yordanova for drawing one figure and are indebted to the reviewer for his critical comments and helpful suggestions that helped us to significantly improve the overall presentation and clarity of the paper.

\medskip
\noindent
\textbf{Disclosure of Potential Conflicts of Interest:} The authors declare that they have no conflicts of interest.
\end{acks}

%
%

\begin{thebibliography}{50}

   \bibitem[\protect\citeauthoryear{Birn \etal}{2005}]{birn2005}
   Birn, J., Galsgaard, K., Hesse, M., Hoshino, M., Huba, J., Lapenta, G., Pritchett, P.L., Schindler, K., Yin, L., B\"{u}chner, J., Neukirch, T., Priest, E.R.: 2005, Forced magnetic reconnection. \emph{Geophys.\ Res.\ Lett.}\ \textbf{32}, L06105. DOI: {\color{blue}10.1029/2004GL022058}. ADS: {\color{blue}http://adsabs.harvard.edu/abs/2005GeoRL..32.6105B}.

   \bibitem[\protect\citeauthoryear{Cally}{1986}]{cally1986}
   Cally, P.S.: 1986, Leaky and non-leaky oscillations in magnetic flux tubes. \emph{Solar Phys.}\ \textbf{103}, 277. DOI: {\color{blue}10.1007/BF00147830}. ADS: {\color{blue}http://adsabs.harvard.edu/abs/1986SoPh..103..277C}.

   \bibitem[\protect\citeauthoryear{Cally and Khomenko}{2015}]{cally2015}
   Cally, P.S., Khomenko, E.: 2015, Fast-to-Alfv\'en Mode Conversion Mediated by the Hall Current. I. Cold Plasma Model. \emph{Astrophys. J.}\ \textbf{814}, 106. DOI: {\color{blue}10.1088/0004-637X/814/2/106}. \\ ADS: {\color{blue}http://adsabs.harvard.edu/abs/2015ApJ...814..106C}.

   \bibitem[\protect\citeauthoryear{Chandrasekhar}{1961}]{chandrasekhar1961}
   Chandrasekhar, S.: 1961, \emph{Hydrodynamic and Hydromagnetic Stability}, Clarendon Press, Oxford, chapter 11.

   \bibitem[\protect\citeauthoryear{Edwin and Roberts}{1983}]{edwin1983}
   Edwin, P.M., Roberts, B.: 1983, Wave propagation in a magnetic cylinder. \emph{Solar Phys.}\ \textbf{88}, 179. DOI: {\color{blue}10.1007/BF00196186}. ADS: {\color{blue}http://adsabs.harvard.edu/abs/1983SoPh...88..179E}.

   \bibitem[\protect\citeauthoryear{Fitzpatrick}{2004}]{fitzpatrick2004}
   Fitzpatrick, R.: 2004, Scaling of forced magnetic reconnection in the Hall-magnetohydrodynamic Taylor problem. \emph{Phys.\ Plasmas\/} \textbf{11}, 937. DOI: {\color{blue}10.1063/1.1768956}. \\ ADS: {\color{blue}http://adsabs.harvard.edu/abs/2004PhPl...11.3961F}.

   \bibitem[\protect\citeauthoryear{Galtier}{2006}]{galtier2006}
   Galtier, S.: 2006, Wave turbulence in incompressible Hall magnetohydrodynamics. \emph{J.\ Plasma Phys.}\ \textbf{72}, 721. DOI: {\color{blue}10.1017/S0022377806004521}. ADS: {\color{blue}http://adsabs.harvard.edu/abs/2006JPlPh..72..721G}.

   \bibitem[\protect\citeauthoryear{Galtier \etal}{2000}]{galtier2000}
   Galtier, S., Nazarenko, S.V., Newell, A.C., Pouquet, A.: 2000, A weak turbulence theory for incompressible magnetohydrodynamics. \emph{J.\ Plasma Phys.}\ \textbf{63}, 447. DOI: {\color{blue}10.1017/S0022377899008284}. ADS: {\color{blue}http://adsabs.harvard.edu/abs/2000JPlPh..63..447G}.

   \bibitem[\protect\citeauthoryear{Ghosh and Goldstein}{1997}]{ghosh1997}
   Ghosh, S., Goldstein, M.L.: 1997, Anisotropy in Hall MHD turbulence due to a mean magnetic field. \emph{J.\ Plasma Phys.}\ \textbf{57}, 129.  DOI: {\color{blue}10.1017/S0022377896005260}. \\ ADS: {\color{blue}http://adsabs.harvard.edu/abs/1997JPlPh..57..129G}.

   \bibitem[\protect\citeauthoryear{Huba}{1995}]{huba1995}
   Huba, J.D.: 1995, Hall magnetohydrodynamics in space and laboratory plasmas. \emph{Phys.\ Plasmas\/} \textbf{2}, 2504. DOI: {\color{blue}10.1063/1.871212}. ADS: {\color{blue}http://adsabs.harvard.edu/abs/1995PhPl....2.2504H}.

   \bibitem[\protect\citeauthoryear{Jones and Downes}{2011}]{jones2011}
   Jones, A.C., Downes, T.P.: 2011, The Kelvin--Helmholtz instability in weakly ionized plasmas: ambipolar-dominated and Hall-dominated flows. \emph{Mon.\ Not.\ Roy.\ Astron.\ Soc.}\ \textbf{418}, 390.  DOI: {\color{blue}10.1111/j.1365-2966.2011.19491.x}. ADS: {\color{blue}http://adsabs.harvard.edu/abs/2011MNRAS.418..390J}.

   \bibitem[\protect\citeauthoryear{Leroy and Keppens}{2017}]{leroy2017}
   Leroy, M.H.J., Keppens, R.: 2017, On the influence of environmental parameters on mixing and reconnection caused by the Kelvin--Helmholtz instability at the magnetopause. \emph{Phys.\ Plasmas\/} \textbf{24}, 012906. DOI: {\color{blue}10.1063/1.4974758}. ADS: {\color{blue}http://adsabs.harvard.edu/abs/2017PhPl...24a2906L}.

   \bibitem[\protect\citeauthoryear{Lighthill}{1960}]{lighthill1960}
   Lighthill, M.J.: 1960, Studies on Magneto-Hydrodynamic Waves and other Anisotropic Wave Motions. \emph{Phil.\ Trans.\ R.\ Soc.\ Lond.}\ A \textbf{252}, 397. DOI: {\color{blue}10.1098/rsta.1960.0010}. \\ ADS: {\color{blue}http://adsabs.harvard.edu/abs/1960RSPTA.252..397L}.

   \bibitem[\protect\citeauthoryear{Mart\'inez-G\'omez \etal}{2015}]{martinez2015}
   Mart\'inez-G\'omez, D., Soler, R., Terradas, J.: 2015, Onset of the Kelvin-Helmholtz instability in partially ionized mafnetic flux tubes. \emph{Astron.\ Astrophys.}\ \textbf{578}, A104.  DOI: {\color{blue}10.1051/0004-6361/201525785}.  ADS: {\color{blue}http://adsabs.harvard.edu/abs/2015A\&A...578A.104M}.

   \bibitem[\protect\citeauthoryear{Nakariakov}{2007}]{nakariakov2007}
   Nakariakov, V.M.: 2007, MHD oscillations in solar and stellar coronae: Current results and perspectives. \emph{Adv.\ Space Res.}\ \textbf{39}, 1804. DOI: {\color{blue}10.1016/j.asr.2007.01.044}. \\ ADS: {\color{blue}http://adsabs.harvard.edu/abs/2007AdSpR..39.1804N}.

   \bibitem[\protect\citeauthoryear{Nykyri and Otto}{2004}]{nykyri2004}
   Nykyri, K., Otto, A.: 2004, Influence of the Hall term on KH instability and reconnection inside KH vortices. \emph{Ann.\ Geophys.}\ \textbf{22}, 935. DOI: {\color{blue}10.5194/angeo-22-935-2004}. \\  ADS: {\color{blue}http://adsabs.harvard.edu/abs/2004AnGeo..22..935N}.

   \bibitem[\protect\citeauthoryear{Panday and Wardle}{2008}]{pandey2008}
   Pandey, B.P., Wardle, M.: 2008, Hall magnetohydrodynamics of partially ionized plasmas. \emph{Mon.\ Not.\ Roy.\ Astron.\ Soc.}\ \textbf{385}, 2269.  DOI: {\color{blue}10.1111/j.1365-2966.2008.12998.x}. \\ ADS: {\color{blue}http://adsabs.harvard.edu/abs/2008MNRAS.385.2269P}.

   \bibitem[\protect\citeauthoryear{Panday}{2013}]{pandey2013}
   Pandey, B.P.: 2013, Surface waves in the partially ionized solar plasma slab. \emph{Mon.\ Not.\ Roy.\ Astron.\ Soc.}\ \textbf{436}, 1659.  DOI: {\color{blue}10.1093/mnras/stt1682}.  ADS: {\color{blue}http://adsabs.harvard.edu/abs/2013MNRAS.436.1659P}.

   \bibitem[\protect\citeauthoryear{Sahraoui \etal}{2007}]{sahraoui2007}
   Sahraoui F., Galtier, S., Belmont, G.: 2007, On waves in incompressible Hall magnetohydrodynamics. \emph{J.\ Plasma Phys.}\ \textbf{73}, 723.  DOI: {\color{blue}10.1017/S0022377806006180}. \\ ADS: {\color{blue}http://adsabs.harvard.edu/abs/2007JPlPh..73..723S}.

   \bibitem[\protect\citeauthoryear{Shadmehri and Downes}{2008}]{shadmehri2008}
   Sadmehri, M., Downes, T.P.: 2008, The role of Kelvin--Helmholtz instability in dusty and partially ionized outflows. \emph{Mon.\ Not.\ Roy.\ Astron.\ Soc.}\ \textbf{387}, 1318.  DOI: {\color{blue}10.1111/j.1365-2966.2008.13345.x}.  ADS: {\color{blue}http://adsabs.harvard.edu/abs/2008MNRAS.387.1318S}.

   \bibitem[\protect\citeauthoryear{Terra-Homem \etal}{2003}]{homem2003}
   Terra-Homem, M., Erd\'elyi, R., Ballai, I.: 2003, Linear and non-linear MHD wave propagation in steady-state magnetic cylinders. \emph{Solar Phys.}\ \textbf{217}, 199.  DOI: {\color{blue}10.1023/B:SOLA.0000006901.22169.59}.  ADS: {\color{blue}http://adsabs.harvard.edu/abs/2003SoPh..217..199T}.

   \bibitem[\protect\citeauthoryear{Zank and Matthaeus}{1993}]{zank1993}
   Zank, G.P., Matthaeus, W.H.: 1993, Nearly incompressible fluids. II: Magnetohydrodynamics, turbulence, and waves. \emph{Phys.\ Fluids\/} \textbf{5}, 257.  DOI: {\color{blue}10.1063/1.858780}. \\ ADS: {\color{blue}http://adsabs.harvard.edu/abs/1993PhFlA...5..257Z}.

   \bibitem[\protect\citeauthoryear{Zaqarashvili \etal}{2014}]{zaqarashvili2014}
   Zaqarashvili, T.V., V\"{o}r\"{o}s, Z., Zhelyazkov, I.: 2014, Kelvin--Helmholtz instability of twisted magnetic flux tubes in the solar wind. \emph{Astron.\ Astrophys.}\ \textbf{561}, A64.  DOI: {\color{blue}10.1051/0004-6361/201322808}.  ADS: {\color{blue}http://adsabs.harvard.edu/abs/2014A\&A...561A..62Z}.

   \bibitem[\protect\citeauthoryear{Zhelyazkov}{2009}]{zhelyazkov2009}
   Zhelyazkov, I.: 2009, MHD waves and instabilities in flowing solar flux-tube plasmas in the framework of Hall magnetohydrodynamics. \emph{Eur.\ Phys.\ J.}\ D \textbf{55}, 127.  DOI: {\color{blue}10.1140/epjd/e2009-00217-3}.  ADS: {\color{blue}http://adsabs.harvard.edu/abs/2009EPJD...55..127Z}.

   \bibitem[\protect\citeauthoryear{Zhelyazkov}{2010}]{zhelyazkov2010}
   Zhelyazkov, I.: 2010, Hall-magnetohydrodynamic waves in flowing ideal incompressible solar-wind plasmas. \emph{Plasma Phys.\ Control.\ Fusion\/} \textbf{52}, 065008.  DOI: {\color{blue}10.1088/0741-3335/52/6/065008}.  ADS: {\color{blue}http://adsabs.harvard.edu/abs/2010PPCF...52f5008Z}.

   \bibitem[\protect\citeauthoryear{Zhelyazkov}{2012}]{zhelyazkov2012}
   Zhelyazkov, I.: 2012, Magnetohydrodynamic waves and their stability status in solar spicules. \emph{Astron.\ Astrophys.}\ \textbf{537}, A124.  DOI: {\color{blue}10.1051/0004-6361/201117780}. \\  ADS: {\color{blue}http://adsabs.harvard.edu/abs/2012A\&A...537A.124Z}.

%
\end{thebibliography}
%

\end{article}
\end{document}